\documentclass[a4paper,11pt]{article}
\pdfoutput=1 

\usepackage{jheppub} 

\usepackage[T1]{fontenc} 

\usepackage{graphicx}	
\usepackage[english]{babel}
\usepackage{amsmath,amsfonts,amsthm,amssymb, graphics, psfrag} 

\usepackage[fleqn,tbtags]{mathtools}

\usepackage{color}

\usepackage[utf8]{inputenc}

\newtheorem{theorem}{Theorem}[section]

\usepackage{amssymb}

\def\be{\begin{equation}}
\def\ee{\end{equation}}
\def\bea{\begin{eqnarray}}
\def\eea{\end{eqnarray}}

\def\cal{\mathcal}

\newcommand{\NN}{{\cal N}}

\newcommand{\la}{\lambda}
\newcommand{\f}{\phi}

\newcommand{\OO}{{\cal O}}
\newcommand{\II}{{\cal I}}
\newcommand{\JJ}{{\cal J}}
\newcommand{\KK}{{\cal K}}

\newcommand{\QQ}{{\cal Q}}
\newcommand{\BB}{{\cal B}}
\newcommand{\EE}{{\cal E}}
\newcommand{\CC}{{\cal C}}

\newcommand{\DD}{{\cal D}}

\global\long\global\long\def\e{\epsilon}

\newcommand{\MM}{{\cal M}}

\newcommand{\Tr}{\mbox{Tr}}
\newcommand{\tr}{\mbox{tr}}

\preprint{DESY-19-212}

\title{\boldmath 4D $\mathcal{N}=2$ SCFTs and spin chains}

\author{Elli Pomoni}

\affiliation{ DESY, Theory Group,
 Notkestrasse 85,  22607 Hamburg, Germany}

\emailAdd{elli.pomoni@desy.de}

\abstract{

\bigskip
This is the writeup of the lectures given at the Winter School ``YRISW 2018''  to appear in a special issue of JPhysA.
In the first part of these lecture notes we review some important facts about 4D  $\mathcal{N}=2$ SCFTs.
We begin with basic textbook material, the supersymmetry algebra and its massless representations and
the construction of
Lagrangians using superspace. Then we turn to  more modern topics, the study of the $\mathcal{N}=2$ SCA and its representation theory.
Our intention is to understand how much we can learn from representation theory alone, even about the dynamics of  $\mathcal{N}=2$ SCFTs.
In the second part of the notes we use these tools to construct spin chains for $\mathcal{N}=2$ SCFTs, the spectral problem of which computes anomalous dimensions of local operators.
We
 discuss their novel features comparing them with their counterparts in $\mathcal{N}=4$ SYM and search for possible integrability structures that emerge. 
}

\begin{document} 
\maketitle
\flushbottom

\newpage

\section{Introduction}

The discovery of integrability in  the planar limit of $\mathcal{N}=4$ SYM   led  to the solution of the spectral problem and is since then being used for the computation of Wilson loops, amplitudes,  correlation functions and other observables. See \cite{Beisert:2010jr} for a review. There is also impressive recent progress on non-planar integrability \cite{Bargheer:2017nne,Bargheer:2018jvq}.
Even though the superconformal symmetry of  $\mathcal{N}=4$ SYM is crucial for these developments, integrability was originally discovered in QCD ($\mathcal{N}=0$ supersymmetry) in the high energy limit of deep inelastic scattering (DIS) \cite{Lipatov:1993yb,Faddeev:1994zg,Korchemsky:1994um}, see \cite{Korchemsky:2010kj} for a review.

A very important open question is which gauge theories are integrable and why, in which limits and which observables can be computed using integrability. With these lectures we will try to  address this question for theories in four dimensions. 
Although it is a very important endeavor  to systematically understand which properties of a gauge theory make it integrable and to classify the gauge theories/observables for which integrability is present, this research direction is very sparsely taken in the literature.
In these lectures we will take a concrete first step in this direction, we will construct spin chains and search for integrability for conformal theories (CFTs) with $\mathcal{N}=2$ supersymmetry, the next simplest class of theories after  $\mathcal{N}=4$ SYM.
Some of the features that we will discover will also remain for certain  $\mathcal{N}=1$ SCFTs.  
Even though we will try to make these lectures pedagogical,  they address  a problem that is not solved in the existing literature and they should be viewed as an invitation for further studies.

\paragraph{Not in these lectures:} The study of $\mathcal{N}=2$ theories is a much broader subject than just the ``AdS/CFT integrability'' direction that we will cover in this lectures. Starting in 1994 with the groundbreaking work of Seiberg-Witten \cite{Seiberg:1994rs} and the microscopic derivation of the instanton partition functions by Nekrasov in 2002 \cite{Nekrasov:2002qd}, 
the activity in the field got reawakened in 2009 with Gaiotto's introduction of class $\mathcal{S}$ of  $\mathcal{N}=2$ CFTs \cite{Gaiotto:2009we}, the AGT correspondence \cite{Alday:2009aq}, the developments on the superconformal index \cite{Rastelli:2014jja} and finally  the $\mathcal{N}=2$ superconformal Bootstrap \cite{Beem:2014zpa} and it's relations to 2D chiral algebras \cite{Beem:2013sza}, as well as the Coulomb Branch operators   \cite{Baggio:2014ioa,Baggio:2014sna,Baggio:2015vxa,Gerchkovitz:2016gxx,Grassi:2019txd}.
It is important to  stress that there is another way integrable models and spin chains appear in the context of  $\mathcal{N}=2$ theories, other than the one we will explore here. Classical integrable systems appear in Seiberg-Witten theory and are associated to the Seiberg-Witten curves \cite{Donagi:1995cf}. These integrable models can be quantised and $q$-deformed by turning on Nekrasov's $\Omega$ background $\epsilon_1,\epsilon_2$. The Nekrasov-Shatashvili limit \cite{Nekrasov:2009rc}, $\epsilon_2 \to 0$, gives a second connection to quantum integrable models and spin chains. The reader interested in these directions is invited to take a look at \cite{Tachikawa:2013kta,Gaiotto:2014bja} and references therein for introductions in some of these other directions.

More than half of these lectures will be devoted to understanding  $\mathcal{N}=2$ theories more generally.
The material that we will present here covers the basic background needed for the study of all the other developments in the field of $\mathcal{N}=2$ theories which we will not include here.
We will begin with the traditional (textbook)  approach to develop the subject that is to study representation theory of the $\mathcal{N}=2$ supersymmetry algebra,
which we will then realise as on-shell massless multiplets in which our fields live. We will describe $\mathcal{N}=2$ theories   using off-shell  $\mathcal{N}=1$ superspace formalism and build Lagrangians imposing the $SU(2)_R$ R-symmetry, which is not manifest in this language. 

In the more modern viewpoint we are instructed to think of  general $\mathcal{N}=2$ theories  as intermediate points of $\mathcal{N}=2$ preserving renormalisation group flows which are starting from an $\mathcal{N}=2$ superconformal UV fixed point  and which may flow to the IR either to a theory with no massless d.o.f., called gapped, or to a CFT. This approach naturally leads to the study of the $\mathcal{N}=2$ superconformal algebra (SCA) and its representation theory which we will pursue in section \ref{sec:SCAreps}.

In section \ref{sec:spinChains} we will turn to our main purpose constructing spin chains for $\mathcal{N}=2$ SCFTs and searching for integrability. We will finish these notes with an overview of the status of the field and possibly interesting open problems, short and long term future goals.

\section*{The spin chain picture}

The spin chain picture is explained in the lectures notes of Marius de Leeuw \cite{deLeeuw:2019usb} in this school/volume.
Here we only present a very quick review so that these notes are self contained and the reader  can go through them smoothly.
The integrability of $\mathcal{N}=4$ SYM, in the planar limit, was first discovered by Minahan and  Zarembo \cite{Minahan:2002ve}.
They showed that the computation of anomalous dimensions (operator mixing) at one-loop can be mapped  to the spectral problem of an integrable spin chain.
In its simplest possible incarnation the problem can be phrased as follows:
  we want to calculate in the large $N$ limit  the anomalous dimension of scalar operators $\mathcal{O}$ that are made out of only two of the three complex scalars, $X,Y,Z$ of $\mathcal{N}=4$ SYM (known as the $SU(2)$ sector),
\begin{equation}
\mathcal{O}=\tr\left(Z^{L-M}X^M \right)  \, .  \nonumber
\end{equation}
Following \cite{Minahan:2002ve} we map this problem to a spin chain by identifying each field 
\be
Z  \quad \longleftrightarrow   \quad \left|   \uparrow \right\rangle      \quad   \mbox{and}   \quad           X   \quad  \longleftrightarrow  \quad\left|   \downarrow\right\rangle    \nonumber \, ,
\ee
with the possible states $ \left|   \uparrow \right\rangle$ and $\left|   \downarrow\right\rangle$ being hosted at a site of a spin chain.
An operator with $L$ constituent fields is mapped to a distribution of spins on a periodic
one-dimensional lattice of length $L$:
\be
\tr \left( ZZZXXZZZXZZZ\ldots\right)  \qquad \longleftrightarrow   \qquad
\left|\uparrow\uparrow\uparrow\downarrow\downarrow
\uparrow\uparrow\uparrow\downarrow
\uparrow\uparrow\uparrow \ldots
\right\rangle
\nonumber \, .
\ee
 The map  is
one-to-one if the spin chain states are required to be translationally
invariant. 

This map is very powerful because,
apart from simply mapping the operators in this sector to spin chain states, we can also  map the anomalous dimensions to energies of the spin chain states.
As we will learn by studying superconformal representation theory, the operators $\tr(Z^L)$ and $\tr(X^L)$ are $\frac{1}{2}$-BPS  and are not allowed by representation theory alone to receive corrections to their conformal dimension. Their anomalous dimension is zero and thus it makes sense to identify them with the  ferromagnetic vacua of the spin chain which have zero energy.

\begin{figure}[h]
   \centering
   \includegraphics[scale=0.8]{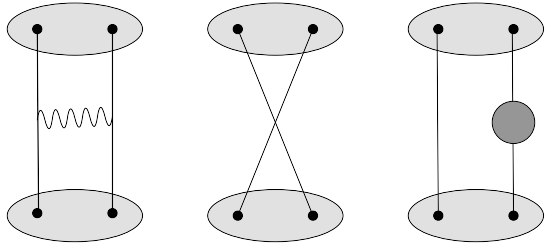} 
   \caption{\it The Feynman diagrams contributing to operator  mixing at one-loop.}
   \label{fig:one-loop}
\end{figure}

The renormalisation  mixing  matrix acts linearly on the operators $\Gamma \, \mathcal{O}_i = \mathcal{Z}_i\,^j  \mathcal{O}_j$ and thus
can be interpreted as a Hamiltonian of a spin chain. It can be computed via Feynman diagrams and at least at one-loop this is a textbook level computation. Due to time constraints we will not discuss how to compute operator renormalisation in perturbation theory in these lectures. The one-loop Feynman diagrams are depicted in Figure \ref{fig:one-loop}. In the large $N$ limit and at the
one-loop level only nearest neighbour interactions are present. 
It turns out that the final result  is equal (up to an overall multiplicative coefficient) to the Hamiltonian of a very well studied spin chain problem,
\begin{equation}
\label{eq:XXX}
\Gamma=\frac{\lambda}{8\pi^2}\sum_{\ell=1}^{L} (  \mathbb{I}-\mathbb{P}_{\ell,\ell+1}) \equiv  \frac{\lambda}{16\pi^2} \mathcal{H}_{XXX}  \, ,
\end{equation}
where $\mathbb{I}$ is the identity operator and $\mathbb{P}_{\ell,\ell+1}$ the permutation operator that permutes spins (states) on neighbour sites $\ell$ and $\ell+1$.
The XXX spin chain is famously integrable and this means that if we know the solution of  the 2-body problem we can get the solution of the $n$-body for free.
Two $X$'s in the sea of $Z$'s are enough to give the anomalous dimension for  $n$ $X$'s in the sea of $Z$'s.

Finally, we would like to end the introduction with the following comment.
The result \eqref{eq:XXX} is due entirely to the $F$-terms, since all other contributions ($D$-terms, gluon exchange
and self-energy diagram) add up to zero. See Section \ref{sec:superspace}.
Some authors refer to this as the ``effective vertex''.
This is an example of a general property of
theories with (extended) supersymmetry and it is known as a non-renormalisation theorem.
In its component form it was discovered in \cite{DHoker:1998vkc,DHoker:2001jzy,Constable:2002hw}.
However, using superspace it becomes powerful \cite{Fiamberti:2008sh,Sieg:2010tz} and makes high-loop computation possible. See \cite{Sieg:2010jt} for a review.

\section{The landscape of 4D  theories}
\label{sec:Landscape}
Before coming to the technical topics, we will begin our lectures with a big picture section. We will review some basic, but very important facts about  4D supersymmetric theories, which should eventually  become clear in the following sections. All possibly unclear or unknown symbols like $\tr \phi^2$, $m \tilde{Q} Q$ or  $\hat{\mathcal{B}}_1$ and $\mathcal{E}_{2}$  for the superconformal multiplets will be explained in Sections \ref{sec:superspace} and \ref{sec:SCAreps}, respectively.
\begin{table}[h]
   \centering
   \begin{tabular}{|c|c|c|}
   \hline
         & vector multiplet & matter multiplet \\
\hline
      $\mathcal{N}=4$   & adj & not possible \\
      $\mathcal{N}=3$   & adj    &  not possible \\
    $\mathcal{N}=2$   & adj  & hypermultiplet in any $\mathcal{R}$ of $G$ \\
     $\mathcal{N}=1$    & adj  & chiral multiplet in any  $\mathcal{R}$  of $G$ \\
      $\mathcal{N}=0$ & adj  & scalar or fermion in any $\mathcal{R}$  of $G$ \\
\hline
   \end{tabular}
   \caption{\it The landscape of  Lagrangian (supersymmetric) theories in four dimensions. 
   For $\mathcal{N} \leq 2$ the matter can be in any representation $\mathcal{R}$ of the gauge group $G$.}
   \label{tab:booktabs}
\end{table}

     $\mathcal{N}=4$ SYM is the maximally supersymmetric theory in four dimensions. It is unique  up to the choice of a semi-simple gauge group $G$.   $\mathcal{N}=4$ supersymmetry is so strong that it leads to further symmetry enhancement, whereby the theory is also conformal. Dynamics are governed by the $psu(2,2|4)$ algebra.\footnote{If we want to preserve  $\mathcal{N}=4$ supersymmetry but are willing to break conformal invariance, it is possible to add a ``$T\bar{T}$ type'' dimension eight irrelevant operator as discussed by Intriligator in \cite{Intriligator:1999ai}.}
      The   $\mathcal{N}=4$  SUSY algebra admits only one possible short massless representation, the ``vector multiplet'' and no ``matter multiplet'' is allowed.
      It is an open problem to prove that  $\mathcal{N}=4$ SYM with color group $G$ is the only possibility with $\mathcal{N}=4$ supersymmetry and explore if  exotic theories exist.

     The next case is  $\mathcal{N}=3$ and enhances to  $\mathcal{N}=4$ SYM  when we demand to have a Lagrangian description. Examples of non-Lagrangian $\mathcal{N}=3$ theories were discovered recently  \cite{Garcia-Etxebarria:2015wns}. However in these lectures we will restrict ourselves to theories with a Lagrangian description, for which we have many more tools to employ.
       
\begin{figure}[t]
\begin{centering}
\includegraphics[scale=0.23]{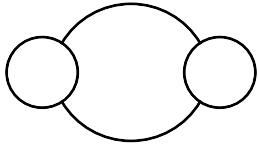}
\par\end{centering}
\caption{\it The quiver of the  $\mathbb{Z}_2$ orbifold of $\mathcal{N}=4$ SYM (the interpolating theory).}
\label{interpolatingquiver}
\end{figure}

\begin{figure}[h]
\begin{centering}
\includegraphics[scale=0.26]{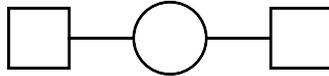}
\par\end{centering}
\caption{\it The quiver diagram of $\mathcal{N}=2$ SuperConformalQCD (SCQCD) with $N_f =2 N$.}
\label{SQCDquiver}
\end{figure}

 Going down to   $\mathcal{N}=2$     supersymmetry we have a huge (largely unexplored and unknown) landscape of theories.
 There exist conformal and non-conformal  theories. For conformal theories with a Lagrangian description we have a complete classification
  by Bhardwaj and Tachikawa \cite{Bhardwaj:2013qia}\footnote{The options are: (i) just a single node of either SU(N), SO(N), USp(N) or G$_2$, (ii) an SU(N) chain, (iii) an SO(N)-USp(N-2) chain and a few exceptional cases.}.
  \begin{theorem}
A Lagrangian theory with $\mathcal{N}=2$  supersymmetry is uniquely specified  via its quiver. It consists of blobs and lines. Blobs correspond to color groups and lines to hypermultiplets. See figures \ref{interpolatingquiver} and \ref{SQCDquiver}. Boxes correspond to global/flavor symmetries.
\end{theorem}
  By now we also know many conformal  theories with no Lagrangian description. They include generalisations of the Argyres-Seiberg \cite{Argyres:2007cn}: $T_N$ trinion theories \cite{Gaiotto:2009we} which can be thought of as  non-Lagrangian generalisation of ``matter multiplets'' in class $\mathcal{S}$, as well as Argyres-Douglas theories \cite{Argyres:1995jj}. This is by no means the complete list.
Starting with conformal theories we can obtain non-conformal theories via triggering an $\mathcal{N}=2$ RG flow with $\mathcal{N}=2$ preserving massive deformations.
 \begin{theorem}
 Possible mass deformations that preserve  $\mathcal{N}=2$  supersymmetry are classified\footnote{Mass deformations are counted by (the Schur or the  Hall-Littlewood limit of)  the SuperConformal Index \cite{Gadde:2011uv} or the Higgs branch Hilbert series. Their number is given by the coefficient of a monomial counting operators with $\Delta=2 R=2$.}. Using $\mathcal{N}=1$ superspace language they can only be of the form $m \tilde{Q} Q$ and according to superconformal representation theory, they are the highest weight state of the $\hat{\mathcal{B}}_1$ superconformal multiplet.
\end{theorem}

 A very good way to obtain many $\mathcal{N}=2$  SCFTs is via orbifolding  $\mathcal{N}=4$ SYM\footnote{For $SU(N)$ color groups we use ADE orbifolds, while for $SO/Sp$ we need to use an orientifold plus an orbifold. Exceptional groups are more complicated to get.}. 
 There is an
ADE classification of $\mathcal{N}=2$  SCFTs with $SU(N)$ color factors using finite/affine Dynkin diagrams. 
The simplest possible example is the $\mathbb{Z}_2$ orbifold of $\mathcal{N}=4$ SYM\footnote{This is a theory with an $AdS_5\times S^5/\mathbb{Z}_2$ gravity dual \cite{Kachru:1998ys,Lawrence:1998ja,Gadde:2009dj} and known to be integrable \cite{Beisert:2005he,Solovyov:2007pw}.
}, with color group  $SU(N)  \times SU(N)$, the quiver of which is depicted in Figure \ref{interpolatingquiver}.
 The orbifolding procedure gives  theories with all the coupling constants equal to each other and to the YM coupling constant of the  $\mathcal{N}=4$ mother theory. This is called the orbifold point. To go away from the orbifold point we marginally deform the theory.
  \begin{theorem}
Marginal operators of $\mathcal{N}=2$  SCFTs are classified:
for theories with a Lagrangian description they are descendants of $\tr  \,\phi^2$ and according to superconformal representation theory they belong to the $\mathcal{E}_{2}$ superconformal multiplet.\footnote{Marginal operators are counted by (the Coulomb limit of)  the SuperConformal Index or the Coulomb branch Hilbert series. Their number is given by the coefficient of a monomial counting operators with $\Delta=r=2$.}
\end{theorem}

Beginning with  the $\mathbb{Z}_2$ orbifold of $\mathcal{N}=4$ SYM and adding a marginal deformation we obtain a
  one parameter family of ${\cal N}=2$ SCFTs with
product gauge group  $SU(N)  \times SU(N)$
and  two exactly marginal couplings $g$ and $\check g$.
In the rest of the lectures we will refer to it as the interpolating theory and we will use it as our basic example. Once spin chains for this example are understood, the generalisation to any Lagrangian $\mathcal{N}=2$ SCFT is straightforward.
\begin{itemize}
\item For $\check g\rightarrow 0$ (ungauging one node) we obtain the  $\mathcal{N}=2$ SuperConformalQCD (SCQCD) with $N_f =2 N$.
\item For  $ \check g= g $ we get back to the
$ \mathbb{Z}_2$ orbifold  of ${\cal N} = 4$ SYM.
\end{itemize}

We finish this section by shortly commenting on $\mathcal{N}=1$ theories.  The $\mathcal{N}=1$ matter multiplets are chiral thus $\mathcal{N}=1$  quivers have arrows contrary to $\mathcal{N}=2$ theories which are not chiral and whose quivers do not have arrows. Generically, their Lagrangians are not completely specified by the quiver. We also have to specify the superpotential(s).
Orbifolding $\mathcal{N}=4$ SYM is a good path to explore the Landscape of $\mathcal{N}=1$ SCFTs which is even more vast and  unexplored, than that of $\mathcal{N}=2$ SCFTs. A lot of the properties  that we will study for the $\mathcal{N}=2$ orbifold daughters of  $\mathcal{N}=4$ SYM  go through for  $\mathcal{N}=1$ orbifold daughters.  A very important class of $\mathcal{N}=1$ orbifold daughters are obtained as $\mathbb{Z}_\ell \times \mathbb{Z}_k$ orbifolds (class $\mathcal{S}_k$ \cite{Gaiotto:2015usa}).

  To understand the statements above we need to study the representation theory of the supersymmetry algebra and then of the superconformal algebra.
 We will see that we can learn a lot just from representation theory, even about dynamics.

 \section{Massless representations of  the supersymmetry algebra}
 \label{sec:masslessReps}

To understand all the facts stated in section \ref{sec:Landscape} we begin with a short review of the supersymmety algebra and its massless representations. This is standard textbook material and more details can be found in any supersymmetry book or review\footnote{The reader who needs ton quickly learn how the detailed calculations can be done can watch the videos from  \href{http://laces.web.cern.ch/Laces/LACES18/videos18.html}{LACES 2018}.}.
 It is worth pointing out that there is a theorem in mathematics (representation theory) which says that there are
 no non-trivial finite dimensional unitary representations
 of non-compact groups.
This impasse was overcome by Wigner with a  trick stemming from his physics intuition. To study representations of the Lorentz/Poincar\'e  group we should go to a reference frame (the rest frame),  classify them there and then boost to  get everything!

The superalgebra generators $\mathcal{Q}_{\alpha}^{A}$ and $\bar{\mathcal{Q}}_{\dot{\alpha}A}$ have spinor indices $\alpha=1,2$ and $\dot\alpha=1,2$ labelling the $SU(2)_{\alpha} \times SU(2)_{\dot\alpha}$ Lorentz group and an additional label $A = 1,2,...,{\cal N}$. The algebra they satisfy is
\bea
\label{eq:susy}
\Big\{\mathcal{Q}_{\alpha}^{A} \ , \ \bar{\mathcal{Q}}_{\dot{\beta}B} \Big\} \, = \, 2 \, (\sigma^{\mu})_{\alpha \dot{\beta}} \, P_{\mu} \, \delta^{A}\,_{B} 
\\
\Big\{\mathcal{Q}_{\alpha}^{A} \ , \ \mathcal{Q}_{\beta}^{B} \Big\}  \, = \, \epsilon_{\alpha \beta} \, Z^{AB}
\label{central}
\eea
with antisymmetric {\em central charges} $Z^{AB} = -Z^{BA}$ commuting with all the generators
and 
\be
\left[\mathcal{Q}_{\alpha}^{A} \ , \ P_\mu \right] = 0 = \left[\bar{\mathcal{Q}}_{\dot{\beta}B} \ , \ P_\mu \right] \, , \qquad \left[ \mathcal{L}_\alpha\,^{\beta}  \ , \  \mathcal{Q}_{\gamma}^{A} \right] = \delta_\gamma\,^{\beta}  \mathcal{Q}_{\alpha}^{A} - \frac{1}{2} \delta_\alpha\,^{\beta}  \mathcal{Q}_{\gamma}^{A}  \, , \quad  \dots
\,.
\ee
The $\mathcal{N} =1$ supersymmetry algebra is invariant under a global phase rotation
of  all supercharges ${Q}_\alpha ^A
$, forming a group $U(1)_r$. 
\be
\left[R \, , \, \mathcal{Q}_{\alpha}^{A}  \right] = + q  \, \mathcal{Q}_{\alpha}^{A}  \qquad \left[R \, , \, \bar{\mathcal{Q}}_{\dot{\beta}B} \right] = - q  \, \bar{\mathcal{Q}}_{\dot{\beta}B} 
\,.
\ee
In extended $\mathcal{N} >1$ supersymmetry algebras, the different supercharges may also be rotated into one
another under the unitary group $SU(\mathcal{N})_R$
\be
\left[R^A\,_B \, , \, \mathcal{Q}_{\alpha}^{C}  \right] = \delta^C\,_B \,  \, \mathcal{Q}_{\alpha}^{A} - \frac{1}{|\mathcal{N}|} \delta^A\,_B \,  \, \mathcal{Q}_{\alpha}^{C} \,.
\ee
 These (outer automorphism) symmetries of the supersymmetry algebra are called
{\bf R-symmetries}. In quantum  field theories, part or all of these
R-symmetries may be broken by anomaly effects.

Concluding our quick presentation of the supersymmetry algebra, it is important to stress that the supersymmetry generators commute with any other bosonic symmetry of the theory like the generators of the color group $G$ or the flavor group $F$,
\be
\label{QTcomutator}
\left[T_G , \, \mathcal{Q}_{\alpha}^{A}  \right] = 0 = \left[T_F , \, \mathcal{Q}_{\alpha}^{A}  \right]  \,.
\ee

\bigskip

We now want to classify one particle states and we only care about massless representations as we want to study only SCFTs.
As the symmetry algebra is non-compact, there are no finite unitary representations. To study them and classify them we go to the rest frame and then boost to get the full particle content.
To get the massless representations of  the supersymmetry algebra we
let the momentum of the state be $p_{\mu} = (E , \ 0 , \ 0 , \ E)$. This is just a convenient choice of frame, a null vector, which commutes with the $SO(2)$ little group (the $L_{12}$ generator of the Lorenz algebra). With this choice, the supersymmetry algebra simplifies to
\be
    \label{susy}
    \Big\{\mathcal{Q}_{\alpha}^{A} \ , \ \bar{\mathcal{Q}}_{\dot{\beta}B} \Big\} =4 \, E \,  \left (   \begin{array}{cc}  1 & 0\\ 0 & 0  \end{array} \right )_{\alpha \dot{\beta}} \delta^{A}\,_{B} 
\ee
which implies that  the algebra for the generator $ \mathcal{Q}_2$ is trivial and will not play any role.  
 The
remaining (active) supercharge operators  $a^A = \frac{1}{2\sqrt{E}}\mathcal{Q}_1^A$ and $(a^A)^\dagger = \frac{1}{2\sqrt{E}} (\mathcal{Q}_1 ^A )^\dagger = \frac{1}{2\sqrt{E}}\bar{\mathcal{Q}}_{\dot 1} ^A$ obey the algebra of $\mathcal{N}$ fermionic creation and annihilation operators
(oscillators). Moreover, their commutator with the helicity generator $L_{12}$ teaches us that they raise and lower helicity as
follows
\begin{itemize}
\item $\left[ L_{12} \, , \, \mathcal{Q}_1 ^A \right] = - \frac{1}{2} \mathcal{Q}_1 ^A $ \   lowers  the  helicity by $1/2$ 
\item $\left[ L_{12} \, , \,\bar{\mathcal{Q}}_{\dot 1} ^A  \right] =  \frac{1}{2}\bar{\mathcal{Q}}_{\dot 1} ^A$ \ raises the
helicity by $1/2$.
\end{itemize}
Defining a Clifford vacuum as the state killed by all $\mathcal{Q}_1 ^A$, we build the massless representation by acting with the helicity raising operators $(\mathcal{Q}_1 ^A)^\dagger$, with $a=1,\cdots , \mathcal{N}$. We label the Clifford vacuum by the helicity $\lambda$.
\be
\nonumber
|\lambda \rangle \longrightarrow  (\mathcal{Q}_1 ^A)^\dagger |\lambda \rangle = |\lambda +\frac{1}{2}\rangle_A   \longrightarrow   (\mathcal{Q}_1 ^A)^\dagger (\mathcal{Q}_1 ^B)^\dagger |\lambda \rangle = |\lambda +1\rangle_{[A,B]}  \longrightarrow \dots  
\qquad  \qquad  \qquad  \qquad
\ee
\be
\nonumber
\qquad  \qquad  \qquad \qquad  \qquad \qquad \qquad  \qquad \qquad  \qquad \dots    \longrightarrow (\mathcal{Q}_1 ^1)^\dagger  \dots  (\mathcal{Q}_1 ^{\mathcal{N}})^\dagger |\lambda \rangle = |\lambda +\frac{\mathcal{N}}{2}\rangle
\ee
Due to the antisymmetry of the R-symmetry  indices $A,B,\dots =1 ,\cdots , \mathcal{N}$, the number of states with helicity $\lambda +\frac{k}{2}$ is $\left(\begin{array}{c} \mathcal{N} \\
     k  \\
   \end{array} 
   \right)$. Thus, the
representation has dimension   (total number of states)
\be
n =
\sum_{k=0}^{\mathcal{N}}   
 \left( \begin{array}{c}     \mathcal{N} \\
     k  \\
   \end{array} 
   \right)  = 2^\mathcal{N} =   2^{\mathcal{N}-1}_{\text{Bosons}} +  2^{\mathcal{N}-1}_{\text{Fermions}} \,.
   \ee  
 The helicity is flipped by CPT, thus to have a physical theory we need to always have both $|\lambda \rangle$ and $|-\lambda \rangle$ states.
We add the CPT conjugate when needed\footnote{Naively the hypermultiplet looks like it is CPT invariant. To recognise that it is not we need to recall that the $SU(2)_R$ doublet of scalars is a pseudo-real representation. For theories with $G=SU(2)$ we could combine $\mathbf{2}_R\times \mathbf{2}_G$ to form a real representation and keep the half-hyper being CPT invariant. For all other cases of groups with higher rank we need to add the CPT conjugate, obtaining the full hypermultiplet.} and obtain:
\be
n=2^\mathcal{N} \times \left\{
\begin{array}{ll} 
1 & \quad\text{if the multiplet is CPT complete }\\
2 & \quad\text{if the CPT conjugate has to be added}
\end{array}
\right.
\ee

\begin{table}[h]
\begin{center}
\begin{tabular}{|c|c| c|c|c|c|c|} \hline
  & $\mathcal{N} =1$ & $\mathcal{N} =1$ & $\mathcal{N} =2$ & $\mathcal{N} =2$ & $\mathcal{N} =3$ & $\mathcal{N} =4$ \\
$\lambda\leq 1$  & vector  & chiral & vector   & hyper  & vector   & vector   \\
\hline \hline
1         & 1      & 0      & 1      & 0      & 1      & 1      \\ \hline
$1/2$     & 1      & 1      & 2      & 1+1      & 3+1    & 4      \\ \hline
0         & 0      & 1+1    & 1+1    & 2+2      & 3+3    & 6      \\ \hline
$-1/2$    & 1      & 1      & 2      & 1+1      & 1+3    & 4      \\ \hline
$-1$      & 1      & 0      & 1      & 0      & 1      & 1      \\ \hline
\hline Total $\#$ & $2 \times 2$ & $2 \times 2$ & $ 2 \times 4$ & $ 2 \times 4$ & $ 2
\times 8$ & 16  \\ \hline
\end{tabular}
\end{center}
\caption{\it The possible  massless supersymmetry multiplets in 4D with numbers of massless states as a function of $\mathcal{N}$ and helicity $\lambda\leq 1$.}
\label{table:masslessSUSYreps}
\end{table}

In Table \ref{table:masslessSUSYreps} we summarize all the possible massless  multiplets of the  4D supersymmetry algebra. 
With  $\mathcal{N} =3$ and $\mathcal{N} =4$ supersymmetry it is not possible to build matter multiplets with helicity $\lambda < 1$.
 The $\mathcal{N} =3$ and $\mathcal{N} =4$ vector multiplets coincide (after the CPT completion for the $\mathcal{N} =3$),
and their quantum field theories are identical (we stress that we demanded that the theory has massless representations). 
The more inexperienced readers should pay attention to the fact that all the fields which we  will use to materialize the content of the
$\mathcal{N} =3$ and $\mathcal{N} =4$ vector multiplets should transform in the same way under the color group $G$, {\it i.e.} in the adjoint representation of $G$.
This  fact stems from \eqref{QTcomutator}.

The $\mathcal{N} =2$ matter multiplets are called hypermultiplets and they are not chiral due to the $SU(2)_R$ R-symmetry of  $\mathcal{N} =2$ .
We see that to specify an $\mathcal{N} =2$ theory we need to choose the color groups $G$ and the representations $\mathcal{R}$ of $G$ of the hypermultiplets. Again, due to
 \eqref{QTcomutator}  all the elements of the hypermultiplet will transform under the same representation $\mathcal{R}$ of $G$ .
All this information we store in a quiver, which consists of blobs and lines. Blobs correspond to color groups and lines to hypermultiplets.
As the  hypermultiplets  are not chiral, the lines do not have arrows. The $\mathcal{N} =1$ matter multiplets are chiral and thus  $\mathcal{N} =1$ quivers have arrows.

 \section{Lagrangians in $\mathcal{N}=1$ superspace language}
 \label{sec:superspace}
 
 To understand the dynamics we need to turn to Lagrangians\footnote{According to the modern Bootstrap approach we don't need to discuss Lagrangians at all and it would be enough to turn to SuperConformal representation theory. However, for the purpose of these lectures,  Lagrangians and superspace are an absolutely necessary tool.} and a very convenient way is to use the $\mathcal{N}=1$ superspace language to construct them.

The $\mathcal{N}=1$ superspace
is an extension of the usual Minkowski spacetime by including  Grassmann (spinor) coordinates\footnote{They anticommute with each other $\theta_\alpha \theta_\beta = -\theta_\beta \theta_\alpha$ but commute with ordinary coordinates 
 $\theta_\alpha x^\mu = x^\mu \theta_\alpha$. Note that they square to zero
 $(\theta_\alpha)^2 = 0$ since $\theta_\alpha \theta_\alpha = -\theta_\alpha \theta_\alpha$.  }
\be
x^\mu \rightarrow  \left( x^\mu  \, , \, \theta_{\alpha}  \, , \, \bar{\theta}_{\dot\alpha} \right)   \qquad \alpha , \dot\alpha=1,2
\ee
As the usual Minkowski coordinates $x^\mu$ are generated by acting with translations $P_\mu = -i \partial_\mu$ on  usual functions $f(x+a)=f(x)+ a^\mu \partial_{\mu}f(x)+\dots$,
 the new  Grassmann coordinates $\theta_{\alpha}$ and $\bar{\theta}_{\dot\alpha}$ are generated by acting with the supersymmetry generators 
$\mathcal{Q}_{\alpha}$ and $\bar{\mathcal{Q}}_{\dot{\beta}}$ of the $\mathcal{N}=1$ supersymmetry algebra on function  $f(x,\theta,\bar{\theta})$.
Due to the  form of the supersymmetry algebra \eqref{eq:susy},
supertranslations (supersymmetry variations)  generated by the action of
\be
\delta f = \left( \xi \mathcal{Q} + \bar{\xi} \bar{\mathcal{Q}}\right) f(x,\theta,\bar{\theta})
\ee
also induce usual translations 
\be
\left( x^\mu  \, , \, \theta_{\alpha}  \, , \, \bar{\theta}_{\dot\alpha} \right)   
 \rightarrow
\left( x^\mu + i\,\theta \sigma^\mu \bar{\xi} - i\,\xi\sigma^\mu \bar{\theta}  \, , \, \theta_{\alpha} + \xi_{\alpha}  \, , \, \bar{\theta}_{\dot\alpha}+ \bar{\xi}_{\dot\alpha}  \right)  \,.
\ee
Thus, the  supersymmetry generators  are represented as differential operators 
\be
\mathcal{Q}_\alpha = \frac{\partial}{\partial \theta^\alpha} - i \, \sigma^\mu_{\alpha \dot\alpha}  \bar{\theta}^{\dot\alpha} \partial_\mu
\, ,
\qquad
\bar{\mathcal{Q}}_{\dot\alpha} = \frac{\partial}{\partial \bar{\theta}^{\dot\alpha} } + i \,{\theta}^{\alpha}  \sigma^\mu_{\alpha \dot\alpha}   \partial_\mu
\, ,
\ee
with their anticommutator defined by the supersymmetry algebra \eqref{eq:susy}.
We also define supercovariant derivatives
\be
D_\alpha = \frac{\partial}{\partial \theta^\alpha} + i \, \sigma^\mu_{\alpha \dot\alpha}  \bar{\theta}^{\dot\alpha} \partial_\mu \, ,
\qquad
\bar{D}_{\dot\alpha} = \frac{\partial}{\partial \bar{\theta}^{\dot\alpha} } - i \,{\theta}^{\alpha}  \sigma^\mu_{\alpha \dot\alpha}   \partial_\mu \, ,
\ee
with the property
\be
 \{ D_{\alpha} \, , \, \bar{D}_{\dot\alpha} \} = 2 \sigma^\mu_{\alpha \dot\alpha} P_\mu 
 \qquad \mbox{and} \qquad   
 \{ D_{\alpha} \, , \, \mathcal{Q}_{\beta} \}  = 
  \{ D_{\alpha} \, , \, \bar{\mathcal{Q}}_{\dot\alpha} \}  = 0 \, .
\ee
Given the fact that the  Grassmann coordinates anticommute, a formal power series expansion in $\theta_{\alpha}$ and $\bar{\theta}_{\dot\alpha}$ terminates.
The most general superfield is 
\be
\Phi(x,\theta,\bar\theta) = \phi(x) + \theta \psi(x) +\bar{\theta} \bar{\chi}(x) + \theta^2 F(x) +  \bar\theta^2 G(x) + \theta\sigma^{\mu}\bar{\theta} A_{\mu}(x)
+ \theta^2 \bar{\theta} \bar{\lambda}(x) +  \bar\theta^2\theta \rho(x) + \theta^2   \bar\theta^2 D(x)
\, .
\ee
This has too many components (degrees of freedom). It corresponds to a reducible representation of the $\mathcal{N}=1$ supersymmetry algebra. To capture the irreducible representations we derived in the previous section we must find ways to  impose  constraints on the superfield such that they commute (anticommute) with the superalgebra.

 We will introduce chiral superfields which are used to describe $\mathcal{N}=1$ matter fields
and vector superfields which will materialise the  $\mathcal{N}=1$ vector multiplets,  which include the  gauge fields.
\begin{table}[h]
\begin{center}
\begin{tabular}{|c||c|c|c||c|c|c|}
\hline
 & $\phi$ & $\psi_\alpha$ & $F$  & $A_\mu$ & $\lambda$ & $D$ \\
\hline
on-shell ($n_B=n_F=2$) & 2 & 2 & 0 & 2 & 2 & 0 \\
\hline
off-shell ($n_B=n_F=4$) & 2 & 4 & 2 & 3 & 4 & 1 \\
\hline
\end{tabular}
\vspace{-0.4cm}
\end{center}
\caption{\it The field content of $\mathcal{N}=1$ Chiral and $\mathcal{N}=1$ Vector multiplets.}
\label{table:onoffshell}
\end{table}
\subsection{Chiral superfields}
By definition a Chiral superfield obeys the constraint
\be
\bar{D}_{\dot\alpha} \Phi =0
\label{eq:Chiral}
\ee
Noting that $\theta$ and $y^{\mu} = x^\mu + i \theta \sigma^\mu \bar{\theta}$ are both annihilated by $\bar{D}_{\dot\alpha}$, it is easy to solve the constraint \eqref{eq:Chiral}.
\bea
&&\Phi(y,\theta) = \phi(y) + \theta \psi(y) + \theta^2 F(y)  =
\\
&& \Phi(x,\theta,\bar\theta)  =  \phi(x) + i \,  \theta\sigma^{\mu}\bar{\theta} \partial_{\mu}\phi(x) + \frac{1}{4}  \theta^2   \bar\theta^2 \Box \phi(x)+ \theta \psi(x) 
- i \,  \theta^2 \sigma^{\mu}\bar{\theta} \partial_{\mu}\psi + \theta^2 F(x)
\nonumber
\eea

\subsection*{Lagrangians for Chiral superfields  and the K\"ahler potential}
A supersymmetry invariant Lagrangian is constructed as the integral of any arbitrary superfield (or combinations of superfields)
\be
\mathcal{L} = \int  d^4\theta \,  \mathcal{K} \left( \Phi^i ,  \bar{\Phi}^{\bar i}  \right) \, , \quad  \mbox{with}  \qquad \delta \mathcal{L} = \partial_\mu(\cdots) \, ,
\ee
where $d^4\theta =  d^2\theta \,  d^2\bar\theta$. What is more, if a term in the Lagrangian is made out of only chiral or only antichiral fields, it is automatically a total derivative
\be
\int  d^4\theta f\left( \Phi \right) =\partial_\mu  \left(\cdots \right)  \, ,
\ee
that means that the theory is invariant up to transformations ({\it K\"ahler transformations})
\be
\label{eq:KahlerTransformations}
\mathcal{K}  \rightarrow \mathcal{K} +   f\left( \Phi \right)+  \bar{f}\left( \bar\Phi \right) \, .
\ee
Let us begin by looking at the most basic combination of superfields  that is supersymmetric, real and has mass dimension four (three properties that our Lagrangian should have)
\be
\label{eq:kahler}
\mathcal{L}_{kin} = \int  d^4\theta \,  \Phi^i   \bar{\Phi}^{\bar i} 
\,  = -  \partial_\mu   \bar{\phi}^{\bar i}  \partial^\mu  \phi^i  + i\, \partial_\mu   \bar{\psi}^{\bar i}  \bar\sigma^\mu  \psi^i  +
 \bar{F}^{\bar i}  F^i  
 \, .
\ee 
Doing so we discovered the kinetic terms of $n$ ($i=1,\dots,n$) scalars, Weyl fermions and auxiliary fields, of $n$  chiral off-shell multiplets (see Table \ref{table:onoffshell}).

We can think of the scalar fields $\phi^i$ as a map from the 4D spacetime to an $n$-complex dimentional target space with complex coordinates $( \phi^i  \, , \,  \bar{\phi}^{\bar i})$. The more general function $\mathcal{K}(\Phi^i ,  \bar{\Phi}^{\bar i})$ is called the {\bf K\"ahler potential} and is a {\it real scalar} function on the target space.
We can use it to define a metric on  the target space ($ds^2 = g_{i \bar{i}}  \,  d\phi^i  \,   d\bar{\phi}^{\bar i}$) and write
\be
g_{i \bar{i}} \equiv \partial_i \partial_{\bar{i}} \mathcal{K} 
\qquad \mbox{and}
\qquad \mathcal{L} = -  g_{i \bar{i}} \partial_\mu   \bar{\phi}^{\bar i}  \partial^\mu  \phi^i  + i\,g_{i \bar{i}}  \partial_\mu   \bar{\psi}^{\bar i}  \bar\sigma^\mu  \psi^i  +
g_{i \bar{i}}  F^i   \bar{F}^{\bar i}  \, .
\ee
The target space because of \eqref{eq:KahlerTransformations} is a K\"ahler manifold and this is entirely due to $\mathcal{N}=1$ supersymmetry.

\subsection*{The Superpotential and F-terms}
The K\"ahler terms give the kinetic terms when $\mathcal{K}$ is quadratic.
 For a more general function $\mathcal{K}$ we obtain extra, non-renormalizable, interaction terms with derivatives. 
For normalisable, non-derivative interaction terms we need holomorphic,  superpotential terms
\be
\label{eq:superpotential} 
\mathcal{L}_{int}  = \int d^2\theta \, \mathcal{W}\left( \Phi\right) + h.c. = F^i  \partial_i \mathcal{W}\left( \phi\right) + \frac{1}{2}  \psi_i  \psi_j    \partial_i  \partial_j \mathcal{W}\left( \phi\right) + h.c. \, ,
\ee
which after integrating out the auxiliary fields
\be
 \bar{F}^{\bar{i}} = -  g^{i\bar{i}} \partial_i \mathcal{W}\left( \phi\right) \quad  \mbox{and} \quad {F}^{{i}} = -  g^{i\bar{i}} \partial_{\bar{i}}  \bar{\mathcal{W}}\left( \bar{\phi}\right)
\ee
lead to
\be
V(\phi) = F^i   \bar{F}^{\bar i}  = \sum_{i}  \left|g^{i\bar{i}} \partial_i \mathcal{W}\left( \phi\right) \right|^2  \geq 0 \,.
\ee
The scalar potential $V(\phi)$ that we will derive now is also a function in this target space.
Note that  the scalar potential in a supersymmetric theory cannot be negative!
A simple but important example is the single chiral multiplet theory with $\mathcal{W}\left( \Phi\right)=\frac{1}{2}m\Phi^2+\frac{1}{3!} \lambda \Phi^3$.
\be
\int d^2\theta \, \mathcal{W}\left( \Phi\right)  = m F \phi +m \psi \psi +\frac{1}{2} \lambda F \phi^2  +  \lambda \psi \psi  \phi  \, ,
\ee
which after integrating out the auxiliary field $\bar{F}=- m  \phi- \frac{1}{2} \lambda  \phi^2$ leads to 
\be
V(\phi) = \bar{F}F =  m^2  \bar{\phi}\phi  +  \frac{1}{4} \lambda^2  (\bar{\phi}\phi )^2 \, .
\ee
Note that the Yukawa $ \lambda \psi \psi  \phi$ and the $\frac{1}{4} \lambda^2  (\bar{\phi}\phi )^2$ coupling are related! This is precisely the reason why in supersymmetric theories miraculous cancellations happen when we compute Feynman diagrams and we never get $\Lambda^2$ divergences (only  $\log \Lambda$).

\subsection{Vector superfield}
By definition the Vector or Real superfield obeys
\bea
&&V= V^\dagger = C + \theta \chi  + \bar\theta \bar\chi  +   \theta^2 \varphi  + \bar\theta^2 \bar\varphi 
+ \theta\sigma^{\mu}\bar{\theta} A_{\mu}
\\
&&+i\, \theta^2 \bar{\theta} \left( \bar{\lambda} + \frac{1}{2}\bar{\sigma}^\mu \partial_\mu \chi \right)  -i\,   \bar\theta^2\theta \left( \lambda  - \frac{1}{2}{\sigma}^\mu \partial_\mu \bar\chi\right) + \frac{1}{2} \theta^2   \bar\theta^2 \left(  D + \Box C \right) \nonumber \, .
\eea 
This superfield still
has too many degrees of freedom ($8_B +8_F$) to  correspond to the massless vector representation of the $\mathcal{N}=1$ supersymmetry algebra. As we will see immediately we will reduce its degrees of freedom via gauge fixing (down to $4_B +4_F$) which after going on shell will be precisely the correct number  ($2_B +2_F$) in table \ref{table:onoffshell}.

The only supersymmetry covariant generalization of the usual $U(1)$ gauge invariance is
\be
V \rightarrow V + i\left(\Lambda -\bar{\Lambda} \right) \, ,
\ee
where $\Lambda$ is a chiral superfield with component expansion
\be
\Lambda = \Lambda + \theta \psi_{\Lambda} +  \theta^2 F_{\Lambda} \, .
\ee
In components the effect of the gauge transformation is
\bea
&& 
\delta C = i\left(\Lambda - \bar\Lambda \right)
\\
&&
\delta \chi = i\, \psi_{\Lambda}  \nonumber
\\
&&
\delta \varphi = i\, F_\Lambda    \nonumber
\\
&&
\delta A_\mu = \partial_\mu  \left( \Lambda + \bar\Lambda \right)
\\
&&
\delta \lambda =0    \nonumber
\\
&&
\delta D = 0    \nonumber  \, .
\eea
The vector field $A_{\mu}$ transforms as usual with {\bf the real part of $\Lambda$}.
The real scalar $C$ transforms with the {\bf imaginary part of  $\Lambda$}.
The abelian gauge symmetry in superspace is larger than the ordinary gauge symmetry.
It is $U(1)_{\mathbb{C}}$ instead of $U(1)_{\mathbb{R}}$.
More generally, the vector superfield is invariant under the complexification $G_{\mathbb{C}}$ of the gauge group  $G$.
The component fields $C, \chi , \varphi$ are {\bf gauge artifacts} and we can gauge them away choosing the ``{\it Wess-Zumino gauge}''.
The WZ gauge fixes the supersymmetric $U(1)_{\mathbb{C}}$ gauge invariance to  $U(1)_{\mathbb{R}}$.
We can slightly relax the WZ gauge to the ``{\it complex gauge}'' and fix only  $\chi=0= \varphi$ and we keep the full complexified $G_{\mathbb{C}}$.

Finally, we just state that  for a non abelian gauge theory  the gauge transformation is
\be
e^{ V} \rightarrow e^{- i \bar\Lambda} e^{V}e^{ i \Lambda}  \, .
\ee

\subsection*{Lagrangians with vector superfields}
To construct a Lagrangian invariant under Lorentz, supersymmetry and gauge transformations it is useful to define the supersymmetric version of the field strength
\be
 W_{\alpha} = -\bar{D}^2( e^{-V} D_\alpha e^V )
\ee
that by construction is a chiral superfield
\be
W_{\alpha}(y,\theta) = \lambda_{\alpha}(y) + \theta^\beta\left( \epsilon_{\beta \alpha} D + F_{\beta \alpha}  \right)(y) + \theta^2 \partial_{\alpha \dot{\alpha} }\bar{\lambda}^{\dot\alpha}(y)
\ee
 and  transforms covariantly
\be
W_{\alpha} \rightarrow e^{-i \Lambda}W_{\alpha} e^{i \Lambda} \, .
\ee
The most obvious real scalar, supersymmetric and gauge invariant combination of $W_{\alpha}$'s with mass dimension four
\be
\mathcal{L} = \int d \theta^2  W^{\alpha}  W_{\alpha}  + h.c. =  -\frac{1}{4} F^{\mu \nu}  F_{\mu \nu} + \frac{1}{4} F^{\mu \nu}  \tilde{F}_{\mu \nu} + \frac{1}{2}  D^2  -    i\,   \bar{\lambda} \mathcal{D}_\mu   \bar\sigma^\mu  \lambda
\ee
is the Lagrangian of $\mathcal{N}=1$ SYM, with
 the covariant derivative $\mathcal{D}_\mu = \partial_\mu + i\,  A_\mu$.

\subsubsection*{Coupling chiral superfields to vector superfields}
To construct the theory of a non abelian gauge group $G$ and some fields $\Phi_i$ in some representation of  $G$,
\be
e^{ V} \rightarrow e^{- i \bar\Lambda} e^{V}e^{ i \Lambda}  \qquad \mbox{and} \qquad \Phi_i \rightarrow e^{-i  \Lambda} \Phi_i   \, ,
\ee
we can write the gauge invariant generalisation of \eqref{eq:kahler} and \eqref{eq:superpotential} 
\be
\mathcal{L} = \int d^4\theta  \,  \Phi^\dagger_i e^{ V} \Phi_i   + \int d^2 \theta \, \mathcal{W}\left( \Phi\right) + h.c.
\ee 
in addition to the $ W^{\alpha}  W_{\alpha}$ part. In the WZ gauge
\bea
\int d^4\theta  \,  \bar{\Phi}_i e^{V} \Phi_i  =  &&  -  \mathcal{D}_\mu   \bar{\phi}^{i}\mathcal{D}^\mu  \phi^i  + i\, \mathcal{D}_\mu   \bar{\psi}^{i}  \bar\sigma^\mu  \psi^i  +
 F^i   \bar{F}^{i} 
 \\
 && - i\,  \bar{\psi}^i \, \bar{\lambda} \, \phi^i + h.c. +   \bar{\phi}^i D \phi^i
 \, .
 \nonumber
\eea

Finally, we can derive the scalar potential $V(\phi) = \bar{F} F + \frac{1}{2}D^2 \geq 0$, as above for chiral fields, and from it obtain the vacuum equations, the $F$- and $D$-flatness conditions
\bea
&& \bar{F}^{i} = -   \partial_i \mathcal{W}\left( \Phi\right) =0  \qquad \forall i
\\
&& D^a =  - \sum_i     \phi_i  \, T^a \,  \bar{\phi}_i  = 0   \nonumber
\eea
where $a=1,\dots , \mbox{rank}(G)$ and $T^a$ are the generators of the Lie algebra $g$.

 \subsection{ $\mathcal{N}=2$ Lagrangians in superspace}

The $\mathcal{N}=2$ {\it  Vector multiplet} contains $(A_\mu \ \lambda _{\alpha}^{\mathcal{I}} \ \phi)
$,  where $\lambda _{\alpha}^{\mathcal{I}}$ are left moving Weyl fermions, and $\phi$ is a
complex scalar. Under $SU(2)_R$ symmetry, $A_\mu$ and $\phi$ are
singlets, while $\lambda^{\mathcal{I}}$ transform as a doublet. Using Witten's diamond diagrams we can depict the field content of an $\mathcal{N}=2$ Vector multiplet as
\be
\begin{array}{ccc}  & A_\mu &
\\  \lambda^1_\alpha&  & \lambda^2_\alpha
\\  & \phi &
\end{array} 
\, ,  \quad    \lambda^{\mathcal{I}}= \left(    \begin{array}{c}  \lambda^1 
\\   \lambda^2
\end{array} \right)    \, ,  \quad  \mathcal{I} =1,2
\nonumber
\ee
where the horizontal axis captures the $SU(2)_R$ symmetry quantum number while the vertical axis captures the $U(1)_r$ symmetry quantum number. In the diamond diagram the diagonals capture  $\mathcal{N}=1$ massless representations
\bea
\mathcal{N}=2 ~ \mbox{Vector} =  ~ ( \mathcal{N}=1 ~ \mbox{Vector} )
  \oplus  ~ ( \mathcal{N}=1 ~ \mbox{chiral}  ) 
\, ,
\eea 
or in terms of $\mathcal{N}=1$ superfields 
$V =\dots + \theta\sigma^{\mu}\bar{\theta} A_{\mu}
 -i\,   \bar\theta^2\theta  \lambda^{1}
 + \dots + \frac{1}{2} \theta^2   \bar\theta^2 D $ and
$ \Phi =  \phi +\theta \lambda^{2}+ \theta^2 F$, both in the adjoint representation  of the gauge group.

\medskip

The $\mathcal{N}=2$ {\it  hypermultiplet} consists of
$(\psi_{\alpha} \ \tilde{\psi}_\alpha\ Q^{\mathcal{I}} )$,
where $\psi_{\alpha}$ and $\tilde{\psi}_\alpha$ are left moving Weyl fermions, while  $Q^{\mathcal{I}}$ are two
complex  scalars, all transforming in some representation ${\cal R}$ of $G$.
Under $SU(2)_R$ symmetry, $\psi$'s are singlets, while $Q^{\mathcal{I}}$
transform as a doublet.
\be
  \begin{array}{ccc}  & \psi_\alpha &
\\   q &  &  \left( \tilde{q} \right)^*
\\  &  \left( \tilde{\psi}_{\alpha}  \right)^\dagger &
\end{array} 
\, ,  \quad    Q^{\mathcal{I}}= \left(    \begin{array}{c} q
\\  \tilde{q}^*
\end{array} \right)  
\, .
\nonumber
\ee
In terms of $\mathcal{N}=1$  massless representations,
\be
\mathcal{N}=2 ~ \mbox{hyper}  ~  = ~ ( \mathcal{N}=1 ~ \mbox{chiral in}  ~ {\cal R} )~  \oplus
~(  \mathcal{N}=1 ~ \mbox{chiral in}  ~ \bar{\cal R} ) 
\,, \qquad \qquad
\ee
\be
Q = q + \theta \psi_{\alpha} +\theta^2 f \quad \mbox{and} \quad
 \tilde{Q} =  \tilde{q} + \theta \tilde{\psi}   + \theta^2 \tilde{f}
\,.
\ee

\medskip

Finally, the $\mathcal{N}=4$ {\it  Vector multiplet} consists of $(A_\mu \ \lambda _\alpha ^A \ X^m)
$, where $\lambda _\alpha ^A$, $A=1,\cdots ,4$ are left moving Weyl fermions
and $X^m$, $m=1,\cdots , 6$ are real scalars. Under $SU(4)_R$
symmetry, $A_\mu$ is a singlet, $\lambda _\alpha ^A$ is a {\bf 4} and the
scalars $X^i$ are an antisymmetric\footnote{${\bf 4} \times {\bf 4} = {\bf 6}_A + {\bf 10}_S$} {\bf 6} $X^{AB}$ representation. 
\bea
\mathcal{N}=4 ~ \mbox{Vector} ~  &=&  ~ \left( \mathcal{N}=2 ~ \mbox{Vector} \right)~  \oplus
~ \left(  \mathcal{N}=2 ~ \mbox{hyper}  \right) ~
\nonumber \\ &=&  ~ \left( \mathcal{N}=1 ~ \mbox{Vector}\right) ~  \oplus
~ 3 \left(  \mathcal{N}=1 ~ \mbox{chiral} \right) \,.
\eea
The  $X^{AB}$  are related to the three complex scalars $\phi^i$ ($i=1,2,3$) in the three chirals
\be
X^{A B} = \left(\begin{array}{cc|cc}
0 & \phi^1 & \phi^2 & \phi^3
\\
-\phi^1 & 0 & \phi^{*}_3 & -\phi^{*}_2
\\
\hline
-\phi^2 & -\phi^{*}_3 & 0 & \phi^{*}_1
\\
-\phi^3 & \phi^{*}_2 &-\phi^{*}_1 & 0
\end{array}\right)
\ee
and obey the self-duality constraint 
 \be \label{selfduality}
(X^{AB})^\dagger  \equiv \bar{X}_{AB} = \frac{1}{2}\epsilon_{ABCD}\,X^{CD} \,.
\ee
The $\mathcal{N} = 4$ vector multiplet splits into
the $\mathcal{N} = 2$ {Vector multiplet} 
\be
\begin{array}{ccc}  & A_\mu &
\\  \lambda^1_\alpha&  & \lambda^2_\alpha
\\  & \phi^1 &
\end{array} \, ,
\ee
and the
 $\mathcal{N} = 2$ {hypermultiplet}
 \be
\begin{array}{ccc}  & \lambda^3_\alpha &
\\     \phi^2  &  &  \phi^3  
\\  & \lambda^4_\alpha &
\end{array}  \,.
\ee 
Using $\mathcal{N} = 1$ language we construct the Lagrangians of   $\mathcal{N}=2$ or  $\mathcal{N}=4$ theories by $(i)$ imposing gauge invariance and $(ii)$  R-symmetry.

\subsubsection*{The $\mathcal{N}=4$ Lagrangian in $\mathcal{N} = 1$ language}
To get $\mathcal{N}=4$ SYM we need to use three chiral superfields $\Phi^i = \phi^i + \theta \psi^i + \theta^2 F^i$ that transform in the adjoint representation. This means that the Lagrangian is
\be
\label{eq:N=4Lagrangian}
\mathcal{L}_{\mathcal{N}=4} = \int d^4\theta  \, \tr \left( e^{-g V}  \Phi^\dagger_i e^{g V} \Phi_i \right) +  \int d \theta^2  \tr \left( W^{\alpha}  W_{\alpha} \right) + \int d^2 \theta \, \mathcal{W}\left( \Phi\right) + h.c.
\, .
\ee 
We will make sure we have $\mathcal{N}=4$ supersymmetry by imposing the $SU({4})$ R-symmetry  on  an $\mathcal{N} = 1$ theory with three chiral and one vector $\mathcal{N}=1$ superfields\footnote{This way of writing is not completely off shell.}.
In  $\mathcal{N} = 1$ language, the  $SU(4)_R$ $R$-symmetry is broken down to an $SU(3) \times U(1)_R$ subgroup.
The $U(1)_R$ is the usual ${\cal N} = 1$ $R$-symmetry,
while the $SU(3)$ is a global symmetry.
  The $SU(3)$ rotates the three chiral superfields leaving $V$ invariant,
 while under the $U(1)_R$ the chiral superfields have  charge $2/3$. 
To derive the Lagrangian all we have to do it to pick the superpotential.
 $SU(4)_R$ $R$-symmetry forbids mass terms.
 The only holomorphic function that we can pick such that it is $SU(3)$ invariant, holomorphic and leads to a renormalizable Lagrangian is
\be
\label{eq:N=4W}
\mathcal{W} = \frac{i\,g}{3!} \epsilon_{ijk}\,\mbox{Tr} \left( \Phi^i \Phi^j \Phi^k  \right)  \, .
\ee
The only ambiguity is the overall coefficient. The way to fix it is by looking at the Yukawa terms: they must appear with the same coefficient so that when we define $\lambda^{A} = (\lambda \ , \ \psi^i)$ we get $ g\,\bar{X}_{AB} \lambda^{A}\lambda^{B}$.
After integrating out the auxiliary fields we get
\begin{displaymath}
{\mathcal L}_{\mathcal{N} = 4} = \mbox{Tr} \Bigg[ -\frac{1}{4}F^{\mu\nu}F_{\mu\nu} + i\bar{\lambda}_{A}\bar{\sigma}^{\mu}D_{\mu}\lambda^{A} - \frac{1}{4}D^{\mu}\bar{X}_{AB} D_{\mu}X^{AB} \Bigg.
 \end{displaymath}
 \be
\Bigg. +i\,\sqrt{2}\, g\,X^{AB}\, \bar{\lambda}_{A}\bar{\lambda}_{B}  -i\,\sqrt{2}\,  g\,\bar{X}_{AB} \lambda^{A}\lambda^{B} - \frac{g^2}{4}[X^{AB}, X^{CD}]\,[\bar{X}_{CD}, \bar{X}_{AB}] \Bigg] \,, 
\ee
where $A, B = 1, \dots,4$.

\subsubsection*{$\mathcal{N}=2$ Lagrangians in $\mathcal{N} = 1$ language}

For theories with $\mathcal{N}=2$ supersymmetry we can obtain the Lagrangian using  $\mathcal{N} = 1$ superspace and imposing the $SU(2)_R$, gauge invariance and  global symmetry invariance. For the vector multiplet the Lagrangian is
\be
\mathcal{L}_{\mathcal{N}=2}^{\text{vec}} = \int d^4\theta  \, \tr \left( e^{-g V}  \Phi^\dagger e^{g V} \Phi \right) +  \int d \theta^2  \tr \left( W^{\alpha}  W_{\alpha} \right)  + h.c. \, .
\ee 
For a fundamental hypermultiplet the Lagrangian is
\be
\mathcal{L}_{\mathcal{N}=2}^{\text{hyper}} = \int d^4\theta  \, \tr \left(  \bar{Q} e^{g V} Q +   \tilde{Q}e^{ - g V}  \bar{\tilde{Q} }    \right) 
+ \, i
 \int d \theta^2  \tr \left( g \tilde{Q} \Phi Q \right) + h.c.  \, ,
 \ee 
while for a bifundamental hypermultiplet
 \bea
\mathcal{L}_{\mathcal{N}=2}^{\text{hyper}} = \int d^4\theta  \, \tr \left( e^{-g_2 V_2}  \bar{Q} e^{g_1 V_1} Q
 + e^{- g_1 V_1}  \bar{\tilde{Q} } e^{g_2 V_2}  \tilde{Q} \right) 
\\
\nonumber \qquad  +\, 
i \int d \theta^2  \tr \left( g_1 \tilde{Q} \Phi_1 Q \right)
- \, 
 i\int d \theta^2  \tr \left( g_2  Q  \Phi_2 \tilde{Q}  \right) + h.c.   \, .
 \eea
 
 \subsubsection*{The vector multiplet in $\mathcal{N}=2$  superspace}
 
 It is desirable to use the formalism in which most of the symmetry is manifest. For the vector multiplet  it is possible to use an off-shell $\mathcal{N}=2$  superspace in which it takes a very simple form. Unfortunately, the hypermultipler in off-shell $\mathcal{N}=2$  superspace is more complicated as we have to use an infinite number of auxiliary fields.
 
 In  real $\mathbb{R}^{4|8}$  superspace with coordinates $\{x,\theta,\tilde\theta  \}$ the  $\mathcal{N}=2$ vector multiplet can be written using the  $\mathcal{N}=2$  chiral superfield strength $\bar{D}_{\dot{\alpha}}\mathcal{W} = 0  =\bar{\tilde{D}}_{\dot{\alpha}}\mathcal{W}$,
 \be
\label{WN2}
\mathcal{W} = \Phi + \tilde\theta^\alpha W_\alpha +  \tilde\theta^2 G \, .
\ee
Within this formalism, the $\mathcal{N}=2$ SYM classical Lagrangian can be compactly written as
\be
\label{classical-prepotential}
\mathcal{L}_{\mathcal{N}=2}^{\text{vec}} =   \int  d^2\theta  d^2\tilde{\theta} \, \tr \big(\mathcal{W}^2\big) = 
 \int d^2\theta  \, \tr \big(W^{\alpha}W_{\alpha}\big)
+  \int  d^2\theta d^2\bar{\theta}  \, \tr \big( e^{-V} \bar{\Phi}e^{V}\Phi \big)  \,.
\ee
This is the formalism Seiberg used to show that the beta function of  $\mathcal{N}=2$  theories is one-loop exact \cite{Seiberg:1988ur}.

\subsection{Lagrangians of orbifold daughters of $\mathcal{N} = 4$ SYM}
\label{subsec:Orbifolds}

\begin{figure}[b]
\begin{centering}
\includegraphics[scale=0.3]{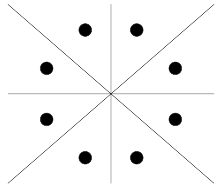}
\par\end{centering}
\caption{\it The $X^6 X^7$ plane in the case of a $\mathbb{Z}_8$ orbifold. Higgsing, which corresponds to placing the $N$ D3 branes away from the origin, allow us to clearly see that $k$ images are created. Open strings with their endpoints on the same stack give rise to the fields of the $\mathcal{N}=2$ vector multiplets. Open strings stretching between different stacks give rise to hypermultiplets. }
\label{fig:Orbifold}
\end{figure}

Consider Type IIB string theory on $\mathbb{R}^{1,3}\times\mathbb{R}^6/\Gamma$  
with $N$ parallel and coincident D$3$ branes along the $\mathbb{R}^{1,3}$, depicted in Table \ref{table:D3Orbifold}. We parametrise the worldvolume of the D$3$ branes with four real coordinates $X^0,X^1,X^2,X^3$, which arrange themselves into the vector representation of the 4D Lorentz group which naturally acts on $\mathbb{R}^{1,3}$. 
The transverse $\mathbb{R}^6$ is parametrised by six real coordinates $X^{4},X^5,X^6,X^7,X^8,X^9$ or alternatively three complex
\begin{equation}
\label{eqn:coordinates}
\phi^1
=\frac{X^5+ i X^6}{\sqrt{2}} \, 
\quad 
\phi^2 =\frac{X^7+ i X^{8}}{\sqrt{2}} \, ,\quad 
\phi^3 =\frac{X^9+ i X^{10}}{\sqrt{2}}
\end{equation}
and their hermitian conjugates.
In the case without an orbifold singularity ($\Gamma = 1$),
 the open strings stretching between D$3$ branes give rise to  $\NN=4$ SYM theory with gauge group  $SU(N)$  on $\mathbb{R}^{1,3}$. The R-symmetry $SU(4)_R  \simeq SO(6)_R$ is identified with the rotations on the transverse $\mathbb{R}^6$. In $\NN=1$ superspace language the theory contains three chiral superfields  $\Phi^1,\Phi^2,\Phi^3$ transforming in the $\mathbf{3}$ of $SU(3)\subset SU(4)_R$,
\begin{equation}
\label{eqn:coordinates}
\phi^1 =\Phi^1|_{\theta=0}  \, ,
\quad 
\phi^2 =\Phi^2|_{\theta=0}  ,\quad 
\phi^3 =\Phi^3|_{\theta=0}   \, .
\end{equation} 
\begin{table}[t]
\centering
\begin{tabular}{ |c |c| c| c| c| c| c| c| c| c| c| }
\hline
   & $X^0$ & $X^1$ & $X^2$ & $X^3$ & $X^4$ & $X^5$ & $X^6$ & $X^7$ & $X^8$& $X^{9}$\\\hline 
 $N$ D$3$ & -- & -- & -- & -- & $\cdot$ & $\cdot$ & $\cdot$ & $\cdot$ & $\cdot$ & $\cdot$\\ \hline
$\mathbb{Z}_{\ell}$ & $\cdot$ & $\cdot$ & $\cdot$ & $\cdot$ & $\cdot$ & $\cdot$ & $\times$ & $\times$ & $\times$ & $\times$\\ \hline
\end{tabular}
\caption{\it The type IIB setup engineering $\NN =2$  orbifold daughters of $\NN =4$ SYM.}
\label{table:D3Orbifold}
\end{table}
The orbifold  $\Gamma=\mathbb{Z}_{\ell}$ acts on the coordinates as
\begin{equation}
\label{eqn:orbifoldactionR10}
\Gamma:\left(\phi^1 , \phi^2  ,\phi^3 \right)\mapsto \left( \phi^1 ,\omega_{\ell} \phi^2 ,\omega_{\ell}^{-1} \phi^3\right) \qquad \mbox{with} \qquad \omega_{\ell}:=e^{2\pi i/\ell} \, ,
\end{equation}
and as the chiral superfields $\Phi^i$ are identified with transverse coordinates \eqref{eqn:coordinates},  the action of $\Gamma$  lies diagonally inside $SU(3)$ in the form
\begin{equation}
R =\begin{pmatrix}
1&0&0\\
0&\omega_{\ell}&0\\
0&0&\omega_{\ell}^{-1}
\end{pmatrix} \in SU(2)_{R} \subset SU(3)\,.
\end{equation}
Note that $\Gamma$ must also have an action inside the gauge group  $SU(N)$ \cite{Douglas:1996sw}. 
To visualise this it is useful to go to the Higgs branch by giving vevs to $\phi^2 = \phi^3 = v$, see  Figure \ref{fig:Orbifold}.
Then, it is clear that for each $D3$ brane, $\ell$ images will be created, as depicted in Figure \ref{fig:Orbifold}.
The action of $\Gamma$  inside the gauge group
 can be conjugated to an element $\tau$ of the maximal torus $T(SU(N))=U(1)^{N-1}$. After scaling $N\to|\Gamma|N=\ell N$ this action breaks $SU(N\ell )\to SU(N)^\ell$.  Hence $\tau$ may be written as
\begin{equation}
\label{eqn:orbifoldactionSUN}
\tau =\mbox{diag} \left(\mathbb{I}, \omega_{\ell}\mathbb{I},\dots, \omega_{\ell}^{\ell-1} \mathbb{I}\right)
\end{equation}
where $\mathbb{I}$ denotes the $N\times N$ identity matrix. Quotienting by $\Gamma$ imposes the identifications
\begin{equation}
\label{eq:orbifoldCondition}
{V}\sim \tau^{\dagger} {V}\tau\,,\quad \Phi^i\sim R^i_{~ j} \tau^{\dagger}\Phi^j \tau \,,
\end{equation}
which implies that the $\ell N \times \ell N$ matrices of $SU(N\ell )$ break in a adjoint of bifundamentals of the $SU(N)^\ell$ quiver 
\begin{eqnarray}
\label{eq:orbifoldConditionExplicit}
V  =\begin{pmatrix}
V_{(1)} &   & & \\
&   V_{(2)} & &\\
& &  & \ddots   & \\
 & & & &  & V_{(k)}  
\end{pmatrix}  
  \,,\quad    
     \Phi^1 =
     \begin{pmatrix}
 & Q_{(1)}  & \\
&  & Q_{(2)} & \\
& &  & \ddots  &  \\
 & & &  &  Q_{(k-1)}  \\
Q_{(k)} &  & & 
\end{pmatrix}  
  \,,\quad    
  \\
     \Phi^2 =  \begin{pmatrix}
  & \tilde{Q}_{(k)} && \\
 \tilde{Q}_{(1)}&  & & \\
& & \ddots  &  \\
 & & &   Q_{(k-1)}  \\
\end{pmatrix}  
\quad \mbox{and} \quad     
      \Phi^3
   =\begin{pmatrix}
\Phi_{(1)} &   & & \\
&   \Phi_{(2)} & &\\
& &  & \ddots   & \\
 & & & &  & \Phi_{(k)}  
\end{pmatrix}  
\, . \nonumber
\end{eqnarray}
After performing these identifications the resulting theory is an $\NN=2$ elliptic quiver gauge theory with gauge group $SU\left(N\right)^{\ell}$ and superpotential
\begin{equation}
\label{eq:orbifoldW}
\mathcal{W}_{\NN=2\, \text{orb.}}
=
i \,g \, \sum_{n=1}^{\ell}  \left( \tilde{Q}_{(n)} \Phi_{(n)} Q_{(n)} -  Q_{(n)} \Phi_{(n+1)}   \tilde{Q}_{(n)}\right) \, ,
\end{equation} 
which is explicitely obtained by plugging \eqref{eq:orbifoldConditionExplicit} in \eqref{eq:N=4W}.
The important point of this construction is that the $\NN=2$ elliptic quiver Lagrangian \eqref{eq:orbifoldW} is the same as the $\NN=4$ one given in  \eqref{eq:N=4W} (and \eqref{eq:N=4Lagrangian}) with the fields obeying the identification \eqref{eq:orbifoldCondition}.

\paragraph{To remember:} The vector multiplet part of $\mathcal{N}=2$ Lagrangians is identical to the  $\mathcal{N}=4$ one.  
For  $\mathcal{N}=2$ orbifold daughters of  $\mathcal{N}=4$  SYM every single vertex is inherited from $\mathcal{N}=4$ - we only need to keep track of the color contractions.

\medskip

The only difference between the $\mathcal{N}=2$ orbifold daughters and their marginal deformaltions is
\begin{equation}
\label{eq:N=2superpotential}
\mathcal{W}_{\NN=2}
=
i\,
\sum_{n=1}^{\ell}  \left(g_{(n)} \tilde{Q}_{(n)} \Phi_{(n)} Q_{(n)} -  g_{(n+1)}  Q_{(n)} \Phi_{(n+1)}   \tilde{Q}_{(n)}\right)  \, .
\end{equation}

\paragraph{The effective vertex:} In computations of anomalous dimensions (especially when they are done in superspace) non-renormalisation theorems help us. Using $\mathcal{N}=1$ superspace it is easy to see that at least to two loops all the work is done by the superpotential \eqref{eq:N=2superpotential}.
We can think of this as an effective vertex that has the structure
 \be
  \label{eq:XXZkappapm}
 \mathbb{I}- \kappa \mathbb{P}_{\ell,\ell+1} \qquad  \mbox{or}  \qquad\mathbb{I}- \kappa^{-1} \mathbb{P}_{\ell,\ell+1}
 \ee
  with $\kappa = g_2/ g_1 $,
$\mathbb{I}$ is the identity and $\mathbb{P}_{\ell,\ell+1}$ the permutation operator that permutes spins (states) on neighbour sites $\ell$ and $\ell+1$ as was the case of $\mathcal{N}=4$  SYM we discussed in the introduction,  equation \eqref{eq:XXX}.

 \section{Basics of representation theory for the $\mathcal{N}=2$ SCA}
 \label{sec:SCAreps}
 
  We begin this section by quickly reviewing some basic facts about the conformal algebra and it's representations so that we can smoothly then turn to the supersymmetric case we are interested in. For a complete and pedagogical review we refer the reader to \cite{Simmons-Duffin:2016gjk}.
We will then turn to the SuperConformal algebra (SCA) with $\mathcal{N}=2$ supersymmetry in 4D and its representations.
We will follow \cite{Dolan:2002zh} where all  the possible shortening  conditions for the $\NN=2$ superconformal algebra were studied and classified.

 \subsection{Conformal Algebra and representations}

 To obtain the conformal algebra, to the Poincar\'e  generators we add the dilatation generator $D$ and the special conformal generator $K_\mu$\footnote{It is useful to know that $K_\mu= I P_\mu I$ with $I$ being an inversion $x^\mu \rightarrow \frac{x^\mu}{x^2}$.}.
 \begin{equation}
\label{eq:ConformalAlgebraKP}
\left[ P_\mu \, , \, K_\nu  \right] = 2 i \left( \delta_{\mu \nu} D + L_{\mu \nu} \right) \, \, , \qquad \left[ D \, , \, P_\mu  \right] = + \, i \, P_\mu 
\, \, , \qquad \left[ D \, , \, K_\mu  \right] = - \, i \, K_\mu
\, .
\end{equation} 
At this stage it is useful to turn from Lorentzian signature to Euclidean and from Minkowski space $\mathbb{R}^{1,3}$ to  $\mathbb{R}^4$. 
It is useful to  think of $\mathbb{R}^4 = \mathbb{S}^3 \times \mathbb{R}$ where the $SO(4) = SU(2) \times SU(2)$ rotations acting on $\mathbb{S}^3$ correspond to spin and the radial translations along $\mathbb{R}$ are generated by the dilatation operator $D$.
 To label states we use their eigenvalues $\Delta , j , \bar{j}$. From \eqref{eq:ConformalAlgebraKP} we see that $P_\mu$ and $K_\mu$ are raising and lowering operators of the conformal dimension $\Delta$. Thus, we construct conformal multiplets by defining a vacuum (the highest weight state) that is annihilated by $K_\mu  |h.w.\rangle = 0$ and acting with $P_\mu$ we obtain the descendants that fill in the multiplet. Note that the conformal multiplets are non-compact and correspond to infinite dimensional representations. 
 
 Using the operator/state correspondence (available in CFTs), the highest weight state $|h.w.\rangle$ corresponds to an operator $\mathcal{O}$ called conformal primary (C.P.). A generic multiplet is labeled by the quantum numbers of the C.P. and is spanned by
  \begin{equation}
\label{eq:ConformalMultiplet}
\mathcal{A}^\Delta_{(j,\bar{j})}  =  \mbox{span} \left\{ P_{\mu_1} \dots P_{\mu_n} \mathcal{O}  \right\}
\end{equation} 
where $ \mathcal{O}$ has conformal dimension $\Delta$ and spin $(j,\bar{j})$ corresponding to the state  $|h.w.\rangle= | \Delta ; j , \bar{j} \rangle$.

To construct unitary representations we impose that the norm of all the states in the multiplet is positive definite $\langle \psi_i | \psi_i \rangle =  || | \psi_i \rangle ||^2 \geq 0$. Starting with the highest weight state $| \psi \rangle = | \Delta ; j , \bar{j} \rangle$, the first level descendants are four states  $P_\mu | \psi \rangle = | \Delta + 1 ; j \pm \frac{1}{2}, \bar{j} \pm \frac{1}{2} \rangle$. Using  \eqref{eq:ConformalAlgebraKP}  we can calculate their norm and find that among them the state with the lowest norm is  $| \psi_1 \rangle = | \Delta + 1 ;  j - \frac{1}{2}, \bar{j} - \frac{1}{2} \rangle$, with norm
  \begin{equation}
\label{eq:LevelOneNorm}
\frac{|| | \psi_1 \rangle ||^2}{2} = \Delta - j -1  + \delta_{j,0} - \bar{j} -1  + \delta_{\bar{j} ,0} =  \Delta - f(j,\bar{j}) \, .
\end{equation} 
Thus, at level one, unitarity imposes
  \begin{eqnarray}
\label{eq:UnitarityLevelOne}
 &\Delta \geq  j + \bar{j} +2&  \qquad \mbox{when both}  \qquad  j   \bar{j} \neq 0    \, ,
 \\
  &\Delta \geq  j +1&  \qquad \mbox{when only}  \qquad  j  \neq 0   \quad \mbox{and} \quad    \bar{j} =0   \, ,
 \\
   &\Delta \geq  \bar{j} +1&  \qquad \mbox{when only}  \qquad  \bar{j}  \neq 0  \quad \mbox{and} \quad    {j} =0    \, ,
   \\
  &\Delta \geq  0&  \qquad \mbox{when both}  \qquad  j = \bar{j} =0   \, .
\end{eqnarray} 
For the C.P. with spin zero we can learn even more by looking for a null vector at level two $P^2 | \psi \rangle = 0$. Using \eqref{eq:ConformalAlgebraKP} we obtain the measure 
$|| P^2 | \psi \rangle || \propto \Delta (\Delta -1)$, which means that for $\Delta \in (0,1)$ the measure is $|| P^2 | \psi \rangle || < 0$. Thus, unitarity demands that $\Delta =0$ or that $\Delta \geq 1$. There is a gap between  $\Delta =0$ and $\Delta  \geq 1$. The operator with   $\Delta =0$ is the identity operator or the vacuum  $| \psi \rangle = |\emptyset\rangle$ of the CFT.
The operator with   $\Delta =1$ obeys  $P^2 | \psi \rangle =0$ and corresponds to a free scalar obeying the equation of motion  $\Box  \phi = 0$. The null vector at level two is the equation of motion  $P^2 | \psi \rangle = | \Box  \phi \rangle  =0$.
When a C.P. saturates the BPS bound $\Delta = f(j,\bar{j})$ the representation is shorter. We have to through away the states with zero measure and all their descendants.

At this stage we can start making statements about the dynamics of CFTs employing representation theory alone.
Short multiplets with  $\Delta = f(j,\bar{j})$ can only acquire anomalous dimensions   $\Delta = f(j,\bar{j}) + \gamma(\lambda)$ (where $\lambda$ is some coupling constant of the theory) {\it only if they can recombine} with some other multiplet to make a long multiplet.
Recombine means that they can gain back the descendants we had to throw away because of the existence of a null vector. 
 Note that the identity operator can never recombine because of the  $\Delta =0$, $\Delta  \geq 1$ gap. In the example of the free scalar,  its multiplet can recombine and  acquire a positive anomalous dimension $\Delta =1 + \gamma_\phi(\lambda) >1$ if there exists an operator $\Box  \phi  \propto \lambda \phi^3 + \dots$ that we can write on the right hand side with $\Delta =3$ at $\lambda =0$  with which $ \Box  \phi$ can mix under renormalisation.  The multiplet of a free scalar is short and is labelled by $\mathcal{B}$. It is obtained from $\mathcal{A}^{\Delta =   1}_{(0,0)}$ after removing the equation of motion $\Box  \phi =0$ and all its descendants (packed in $\mathcal{A}^{\Delta =   3}_{(0,0)}$).

\begin{table}[h]
{\small
\begin{centering}
 \renewcommand*{\arraystretch}{1.4}
\begin{tabular}{|c|l|l|l|l|}
\hline 
\multicolumn{3}{|c|}{Shortening Conditions} & Multiplet\tabularnewline
\hline
\hline 
$\CC$   &  & $\Delta=j+ \bar{j}+2$  & $\CC_{(j,\bar{j})}$\tabularnewline
\hline 
$\BB_R$   & $j=0$ & $\Delta=\bar{j}+1$  & $\BB_{(0,\bar{j})}$\tabularnewline
\hline 
$\BB_L$   & $\bar j=0$ & $\Delta=j+1$  & $\bar{\BB}_{(j,0)}$\tabularnewline
\hline 
 $\BB$ &      $j = \bar j=0$  & $\Delta=1$  & $\BB$\tabularnewline
\hline
\end{tabular}
\par\end{centering}
}
\caption{\it Shortening conditions 
and short multiplets for the conformal algebra.
\label{table:conformalShortening}
} 
\end{table}

At this stage it is better to proceed with more examples. A second familiar case is a free fermion with $\Delta  = j + 1 = \frac{1}{2} + 1$ for which the null vector $P^{\alpha \dot\alpha}   | \psi_\alpha \rangle = \partial^{\alpha \dot\alpha}  \psi_\alpha =0$ corresponds to the free Weyl equation of motion, or similarly $P^{\alpha \dot\alpha}   |\bar{\psi}_{\dot\alpha} \rangle = \partial^{\alpha \dot\alpha}  \bar{\psi}_{\dot\alpha} =0$. These multiplets are labeled as $\mathcal{B}_{(1/2,0)}$ and  $\mathcal{B}_{(0,1/2) }$ respectively and they can only recombine if there is an operator with dimension $5/2$ and spin $(0,1/2)$ or $(1/2,0)$ such that
$\partial^{\alpha \dot\alpha}  \bar{\psi}_{\dot\alpha} \propto \lambda \phi \psi^\alpha + \dots$ to fill in the place of the null vector.

A third important example is a conserved current $J_\mu$ that corresponds to the state $ |3; 1/2 , 1/2 \rangle$  with $\Delta =  j + \bar{j} +2 = 1/2 + 1/2 + 2 = 3$ as in \eqref{eq:UnitarityLevelOne}. The null vector $P |3 ; 1/2 , 1/2 \rangle =  |4; 0 , 0 \rangle = \partial^\mu J_\mu = 0$ corresponds to the conservation law of the current. In a CFT if there exists an operator $J_\mu$ with $\Delta =3$, it is automatically a conserved current and vice versa: if a vector field is conserved it has $\Delta =3$. Conservation implies absence of anomalous dimensions. These shorter representations are denoted as  $\mathcal{C}_{(j,\bar{j})}= \mathcal{C}^{\Delta = j + \bar{j} +2 }_{(j,\bar{j})}$ with  $\mathcal{C}_{(1/2,1/2)}$ being the multiplet of a conserved current.

Finally,  the stress energy tensor  is the primary for the  $\mathcal{C}_{(1,1)}$ multiplet.  $T_{\mu \nu}$ corresponds to a state with  $|4 ; 1,1 \rangle$ and the null vector  $P |4 ; 1 , 1 \rangle =  |5; 1/2 , 1/2 \rangle = \partial^\mu T_{\mu \nu} = 0$ corresponds to the conservation of the stress energy tensor.

Combining everything we learned above, the only way for an operator to obtain an anomalous dimensions $\Delta = \Delta_{0} + \gamma(\lambda)$ is to recombine. The only way this can be done is if there is another multiplet in the CFT that has conformal dimension and spin which are the same as those of the null vector we had to through away. All the possible ways this can happen are summarised by the {\it recombination rules}, 
\bea
\label{eq:RecombinationRules1}
&& \lim_{\gamma \to 0} \left[ \mathcal{A}^{\Delta =   j + \bar{j} +2+ \gamma }_{(j,\bar{j})}  \right] 
=  \mathcal{C}_{(j,\bar{j})}  \oplus  \mathcal{A}^{\Delta =   j + \bar{j} +3 }_{(j - \frac{1}{2},\bar{j}- \frac{1}{2})} 
\\
\label{eq:RecombinationRules2}
&&  \lim_{\gamma \to 0} \left[ \mathcal{A}^{\Delta =   j + 1 + \gamma }_{(j,\bar{j})}  \right] 
= \mathcal{B}^L_j   \oplus   \mathcal{C}_{(j - \frac{1}{2} , \frac{1}{2})} 
\\
\label{eq:RecombinationRules3}
&&  \lim_{\gamma \to 0} \left[ \mathcal{A}^{\Delta =   \bar{j}  + 1 + \gamma }_{(j,\bar{j})}  \right] 
= \mathcal{B}^R_{\bar{j} }   \oplus   \mathcal{C}_{( \frac{1}{2}, j - \frac{1}{2} )} 
\\
\label{eq:RecombinationRules4}
&&  \lim_{\gamma \to 0} \left[ \mathcal{A}^{\Delta =   1 + \gamma }_{(0,0)}  \right] 
= \mathcal{B}  \oplus   \mathcal{A}^{\Delta =  3 }_{(0,0)}  \, .
\eea
Finding ways to explicitly apply these recombination rules to certain CFTs can lead to  very impressive results unveiling the dynamical of the theory \cite{Rychkov:2015naa}.

 \subsection{SuperConformal Algebra and representations}
 
We begin by recalling that in Lorentzian signature $\bar{\mathcal{Q}} = {\mathcal{Q}}^\dagger$ is the complex conjugate of $\mathcal{Q}$. In Euclidian they are independent and maybe $\tilde{\mathcal{Q}}$ is a better notation.
 To write down the  superconformal algebra (SCA) in 4D we need for each $\mathcal{Q}_{\alpha}^{A}$ to introduce its $\mathcal{S}^{\alpha}_A$
 (in radial quantization $\mathcal{S}=\mathcal{Q}^\dagger$ the same way $K = P^\dagger$)
 \bea
&\Big\{\mathcal{Q}_{\alpha}^{A} \ , \ \tilde{\mathcal{Q}}_{\dot{\beta}B} \Big\} \, = \, 2 \, P_{\alpha \dot{\beta}}  \, \delta^{A}\,_{B} 
\qquad
\Big\{\mathcal{S}^{\alpha}_{A} \ , \ \tilde{\mathcal{S}}^{\dot{\beta}B} \Big\} \, = \, 2 \, K^{\alpha \dot{\beta}}  \, \delta_{A}\,^{B} 
&
\\
&
\Big\{\mathcal{Q}_{\alpha}^{A} \ , \ \mathcal{Q}_{\beta}^{B} \Big\}  \, = 0
\qquad
\Big\{\mathcal{S}^{\alpha}_{A} \ , \ \mathcal{S}^{\beta}_{B} \Big\}  \, = 0
&
\label{central}
\eea
and most importantly\footnote{To derive the precise factors of 2 and signs you need to check Jacobi identities.}
\be
\Big\{\mathcal{Q}_{\alpha}^{A} \ , \ {\mathcal{S}}_{B}^\beta \Big\} \, = \, 4 \left( \delta^{A}\,_{B}  \left(\mathcal{L}_\alpha\,^\beta - \frac{1}{2} \delta_\alpha\,^\beta  D\right)  - \delta_\alpha\,^\beta R^{A}\,_{B} \right)
\,,
\ee
while
\be
\Big\{\mathcal{Q}_{\alpha}^{A} \ , \ \tilde{\mathcal{S}}_{B \dot{\beta}} \Big\} \, = 0 = \, \Big\{\tilde{\mathcal{Q}}_{A \dot{\alpha}} \ , \ {\mathcal{S}}_{B}^\beta \Big\} 
\, 
\ee
and 
\be
\label{eq:DQ}
\left[ D \, , \, \mathcal{Q}_{\alpha}^{A}\right] = + \frac{1}{2}  \mathcal{Q}_{\alpha}^{A} 
\quad , \qquad
\left[ D \, , \,  \tilde{\mathcal{Q}}_{\dot{\alpha}A} \right] = + \frac{1}{2}  \tilde{\mathcal{Q}}_{\dot{\alpha}A} 
\ee
\be
\label{eq:DS}
\left[ D \, , \, \mathcal{S}^{\alpha}_{A}\right] = - \frac{1}{2}   \mathcal{S}^{\alpha}_{A}
\quad , \qquad
\left[ D \, , \, \tilde{\mathcal{S}}^{\dot{\alpha}A}\right] = - \frac{1}{2}  \tilde{\mathcal{S}}^{\dot{\alpha}A}
\, .
\ee

Following what we previously learned about the conformal representation theory,
 from \eqref{eq:DQ} and \eqref{eq:DS} we see that $\mathcal{Q}$ and $\mathcal{S}$ raise and lower the conformal dimension by $1/2$, respectively. A  SuperConformal primary (S.C.P.) is by definition 
 annihilated by all  $4\mathcal{N}=8$ conformal supercharges $\mathcal{S}^{\alpha}_{A}$ and $\tilde{\mathcal{S}}^{\dot{\alpha}A}$. A superconformal multiplet is generated by the action of the $4\mathcal{N}=8$ Poincar\'e supercharges
$\QQ$ and $\tilde {\QQ}$ on the S.C.P..
A generic long multiplet  of the $\NN=2$
SCA  is labeled by the quantum numbers of its S.C.P., the eigenvalues $(\Delta, R, r, j , \bar j)$ of the Dilatation operator,  the Cartan
generators of the $SU(2)_{R} \times U(1)_{r}$ R-symmetry  and of the Lorentz group, and is denoted by $\mathcal{A}_{R,r(j,\bar{j})}^{\Delta}$. 
 Note that $\mathcal{S} ,\tilde{ \mathcal{S}}|h.w.\rangle = 0$ implies that  $K |h.w.\rangle = 0$ so a S.C.P. is also a primary of the conformal algebra.
 Moreover, to construct the multiplets we can act with  $\QQ$'s either in a symmetrized way which creates $P$ and thus conformal primaries, or
 with anti-symmetrized action which keeps us in the finite superconformal multiplet.
 When  some combination of
the  $\QQ$'s  also annihilates the primary, the corresponding multiplet
is shorter. 
$|R,r\rangle^{h.w.}_{(j,\bar{j})}$ is the highest weight state with eigenvalues $(R, r, j , \bar j)$ under the Cartan
generators of the $SU(2)_{R} \times U(1)_{r}$ R-symmetry  and of the Lorentz group.
The multiplet  built on this state is  denoted as $\mathcal{X}_{R,r(j,\bar{j})}$,
where the letter $\mathcal{X}$ characterizes the shortening condition.
The left column of Table \ref{shortening} labels
the condition. 
Note that conjugation reverses the signs of $r$, $j$ and $\bar j$ in the expression of the conformal dimension.

\medskip

\begin{table}[h!]
{\small
\begin{centering}
 \renewcommand*{\arraystretch}{1.4}
\begin{tabular}{|c|l|l|l|l|}
\hline 
\multicolumn{4}{|c|}{Shortening Conditions} & Multiplet\tabularnewline
\hline
\hline 
$\BB_{1}$  & $\QQ_{\alpha}^{1}|R,r\rangle^{h.w.}=0$  & $j=0$ & $\Delta=2R+r$  & $\BB_{R,r(0,\bar{j})}$\tabularnewline
\hline 
$\bar{\BB}_{2}$  & $\tilde{\QQ}_{2 \dot{\alpha}}|R,r\rangle^{h.w.}=0$  & $\bar j=0$ & $\Delta=2R-r$  & $\bar{\BB}_{R,r(j,0)}$\tabularnewline
\hline 
$\EE$  & $\BB_{1}\cap\BB_{2}$  & $R=0$  & $\Delta=r$  & $\EE_{r(0,\bar{j})}$\tabularnewline
\hline 
$\bar \EE$  & $\bar \BB_{1}\cap \bar \BB_{2}$  & $R=0$  & $\Delta=-r$  & $\bar \EE_{r(j,0)}$\tabularnewline
\hline 
$\hat{\BB}$  & $\BB_{1}\cap\bar{B}_{2}$  & $r=0$, $j,\bar{j}=0$  & $\Delta=2R$  & $\hat{\BB}_{R}$\tabularnewline
\hline
\hline 
$\CC_{1}$  & $\e^{\alpha\beta}\QQ_{\beta}^{1}|R,r\rangle_{\alpha}^{h.w.}=0$  &  & $\Delta=2+2j+2R+r$  & $\CC_{R,r(j,\bar{j})}$\tabularnewline
 & $(\QQ^{1})^{2}|R,r\rangle^{h.w.}=0$ for $j=0$  &  & $\Delta=2+2R+r$  & $\CC_{R,r(0,\bar{j})}$\tabularnewline
\hline 
$\bar \CC_{2}$  & $\e^{\dot\alpha\dot\beta}\tilde\QQ_{2\dot\beta}|R,r\rangle_{\dot\alpha}^{h.w.}=0$  &  & $\Delta=2+2\bar j+2R-r$  & $\bar\CC_{R,r(j,\bar{j})}$\tabularnewline
 & $(\tilde\QQ_{2})^{2}|R,r\rangle^{h.w.}=0$ for $\bar j=0$  &  & $\Delta=2+2R-r$  & $\bar\CC_{R,r(j,0)}$\tabularnewline
\hline 
$\mathcal{F}$  & $\CC_{1}\cap\CC_{2}$  & $R=0$  & $\Delta=2+2j+r$  & $\CC_{0,r(j,\bar{j})}$\tabularnewline
\hline 
$\bar{\mathcal{F}}$  & $\bar\CC_{1}\cap\bar\CC_{2}$  & $R=0$  & $\Delta=2+2\bar j-r$  & $\bar\CC_{0,r(j,\bar{j})}$\tabularnewline
\hline 
$\hat{\CC}$  & $\CC_{1}\cap\bar{\CC}_{2}$  & $r=\bar{j}-j$  & $\Delta=2+2R+j+\bar{j}$  & $\hat{\CC}_{R(j,\bar{j})}$\tabularnewline
\hline 
$\hat{\mathcal{F}}$  & $\CC_{1}\cap\CC_{2}\cap\bar{\CC}_{1}\cap\bar{\CC}_{2}$  & $R=0, r=\bar{j}-j$ & $\Delta=2+j+\bar{j}$  & $\hat{\CC}_{0(j,\bar{j})}$\tabularnewline
\hline
\hline 
$\DD$  & $\BB_{1}\cap\bar{\CC_{2}}$  & $r=\bar{j}+1$  & $\Delta=1+2R+\bar{j}$  & $\DD_{R(0,\bar{j})}$\tabularnewline
\hline 
$\bar\DD$  & $\bar\BB_{2}\cap{\CC_{1}}$  & $-r=j+1$  & $\Delta=1+2R+j$  & $\bar\DD_{R(j,0)}$\tabularnewline
\hline 
$\mathcal{G}$  & $\EE\cap\bar{\CC_{2}}$  & $r=\bar{j}+1,R=0$  & $\Delta=r=1+\bar{j}$  & $\DD_{0(0,\bar{j})}$\tabularnewline
\hline
$\bar{\mathcal{G}}$  & $\bar\EE\cap{\CC_{1}}$  & $-r=j+1,R=0$  & $\Delta=-r=1+j$  & $\bar\DD_{0(j,0)}$\tabularnewline
\hline
\end{tabular}
\par\end{centering}
}
\caption{\it Shortening conditions 
and short multiplets for the  $\NN=2$ superconformal algebra.
\label{shortening}
} 
\end{table}

There are three basic types of shortening ($\mathcal{A}$, $\BB$ and $\CC$):
\begin{itemize}
\item  $\mathcal{A}$-type: No shortening condition:  $\mathcal{A}_{R,r(j,\bar{j})}^{\Delta}$ generic long multiplet  of the $\NN=2$ SCA.
\item  $\mathcal{B}$-type: $\QQ_{\alpha} |R,r\rangle^{h.w.} =0$ for both $\alpha = +$ and $-$ which means that 2 supercharges kill the highest weight state and is only possible when the highest weight  has $j=0$. 
 (or $\bar{\mathcal{B}}$: $\tilde{\QQ}_{\dot\alpha} |R,r\rangle^{h.w.} =0$  only possible when $\bar{j}=0$)
 \\
 For $\mathcal{N}=2$ we have two  $\mathcal{B}$-type conditions:
\begin{itemize}
\item   $\mathcal{B}^{1}$: $\QQ_{\alpha}^{1} |R,r\rangle^{h.w.} =0$ (or $\bar{\mathcal{B}}_{1}$: $\tilde{\QQ}_{1 \, \dot{\alpha}} |R,r\rangle^{h.w.} =0$)
\item  $\mathcal{B}^{2}$: $\QQ_{\alpha}^{2} |R,r\rangle^{h.w.} =0$ (or $\bar{\mathcal{B}}_{2}$: $\tilde{\QQ}_{2 \, \dot{\alpha}} |R,r\rangle^{h.w.} =0$).
\end{itemize}
This type of shortening is $\frac{1}{4}$-BPS (two out of eight  $\QQ$s).
\item  $\mathcal{C}$-type:  $\e^{\alpha\beta}\QQ_{\beta}|R,r\rangle_{\alpha}^{h.w.}=0$ which means that only one (combination of) supercharge(s) kills the highest weight state. This condition is half as strong as the  $\mathcal{B}$-type.
 \\
 For $\mathcal{N}=2$ we have two  $\mathcal{C}$-type conditions:
\begin{itemize}
\item   $\mathcal{C}^{1}$: $\e^{\alpha\beta}\QQ_{\beta}^{1} |R,r\rangle^{h.w.} =0$ (or $\bar{\mathcal{C}}_{1}$)
\item  $\mathcal{C}^{2}$: $\e^{\alpha\beta}\QQ_{\beta}^{2} |R,r\rangle^{h.w.} =0$ (or $\bar{\mathcal{C}}_{2}$).
\end{itemize}
 $\mathcal{C}$-type shortening is $\frac{1}{8}$-BPS (one out of eight  $\QQ$s).
\end{itemize}
Now we can list all the possible combinations of the shortening conditions above. This was done by \cite{Dolan:2002zh}  and we
 summarise their findings in Table \ref{shortening}.
There are three types of multiplets that are $\frac{1}{2}$-BPS. We have in total
$\mathcal{Q}_{\alpha}^{A}$, $\tilde{\mathcal{Q}}_{\dot{\beta}B}$: $2\times2+2\times2=8$ $Q$s. $\frac{1}{2}$-BPS means that 4 supercharges kill the primary.
For these lectures it is important for the reader to at least  pay attention to these maximally short, irreducible superconformal representations:

\begin{itemize}
\item  $\EE_{r}= \EE_{r(0,0)}$   whose highest weight state  is materialised by $\tr \bar{\phi}^\ell$ with $\ell = r$ (see   Table  \ref{table:SQCDquantumnumbers} for our convensions) and obeys the shortening condition $\Delta = r $ from $\tilde{\mathcal{Q}}_{\II \, \dot{\alpha}} |h.w.\rangle=0$ for all $\II$,$\dot{\alpha}$. The scalar $\phi$ is the adjoint complex scalar inside the vector multiplet. 
\[
\begin{array}{l|cccccc}
\Delta\\
\ell & 0_{\left(0,0\right)}\\
\ell+\frac{1}{2} &  & \frac{1}{2}_{\left(0,\frac{1}{2}\right)}\\
\ell+1 &  &  & 0_{\left(0,1\right)}, {1_{\left(0,0\right)}}\\
\ell+\frac{3}{2} &  &  &  & \frac{1}{2}_{\left(0,\frac{1}{2}\right)}\\
\ell+2 &  &  &  &  &  & 0_{\left(0,0\right)}\\
\hline
r & \ell \quad& \ell-\frac{1}{2} & \ell-1 & \ell-\frac{3}{2} &  & \quad \ell-2\end{array}
\]
The case $\EE_{1}$ is special as it is the $\mathcal{N}=2$ vector multiplet we discovered above studying the supersymmetry algebra (together with its equations of motion   and the auxiliary field). The field content of the $\mathcal{N}=2$ vector multiplet without the equations of motion (after they are removed as they correspond to null vectors) is captured by  $\DD_{0(0,0)}$.
The case $\EE_{2}$ is also important as it contains the Lagrangian of  $\mathcal{N}=2$ theories as a descendant. Schematically, the Lagrangian is $\mathcal{L}=Q^4 \tr \phi^2$.
The highest weight operators of  $\EE_{r}$ parameterise the {\bf Coulomb branch} (= supersymmetric vacua with $\langle \phi \rangle = a$ and $\langle Q \rangle = 0$, important for other applications).

\item  $\hat{\BB}_{R}$   whose higherst weight state obeys  $\Delta = 2 R $. This shortening condition requires $r=0$, $j=\bar{j}=0$. 
\[ \label{B1multiplet}
\begin{array}{l|ccccc}
\Delta\\
2 &  &  &  {1_{\left(0,0\right)}}\\
\frac{5}{2} &  &  {\frac{1}{2}_{\left(\frac{1}{2},0\right)}} &  & \frac{1}{2}_{\left(0,\frac{1}{2}\right)}\\
3 & 0_{\left(0,0\right)} &  & 0_{\left(\frac{1}{2},\frac{1}{2}\right)} &  & 0_{\left(0,0\right)}\\
\frac{7}{2} &&&\\
4 &  &  & -0_{\left(0,0\right)}\\
\hline
r & 1 & \frac{1}{2} & 0 & -\frac{1}{2} & -1\end{array}
\]
For $R=1$, $\Delta=2$ the  $\hat{\BB}_{1}$ multiplet has as its highest weight state the mesonic operator $\MM_{\mathbf{3}}$ of the $\mathcal{N}=2$ chiral ring (a.k.a. moment map) which is a triplet of the $SU(2)_R$. It also  contains the flavor current as the vector field with $\Delta = 3$ and labeled by $0_{\left(\frac{1}{2},\frac{1}{2}\right)}$. 
The $\Delta = 4$ element denoted as $-0_{\left(0,0\right)}$ corresponds to the conservation of the flavor current.
For $R=1/2$ we get the hypermultiplet. 
The highest weight operators of $\hat{\BB}_{R}$  parameterise the {\bf Higgs branch} (= susy vacua with $\langle \phi \rangle = 0$ and $\langle Q \rangle \neq 0$).

\item  $\hat{\CC}_{0 (j,\bar{j})} = \hat{\CC}_{(j,\bar{j})}$ with shortening condition $\Delta = 2+ j +\bar{j}$. For $j = \bar{j}=0$ the multiplet $\hat{\CC}_{0 (0,0)}$ {\bf contains the stress energy tensor} (the state $0_{\left( 1,1 \right)}$ with $\Delta=4$), the supercurrents  (the states $\frac{1}{2}_{\left( 1, \frac{1}{2} \right)}$ and $\frac{1}{2}_{\left(  \frac{1}{2},1 \right)}$  with $\Delta=7/2$) and $SU(2)_{R}$ and the $U(1)_{r}$ R-symmetry currents ($1_{(\frac{1}{2},\frac{1}{2})}$ and $0_{\left(  \frac{1}{2}, \frac{1}{2} \right)}$ respecively) of the $\NN=2$ theory. 
\begin{equation}
    \begin{array}{l|ccccc}
 \Delta      &                                                                              \\
 &&&&\\
 2          &   &  &0_{\left( 0,0 \right)}                                         &\\
 &&&&\\
 \frac{5}{2} &    &       { \frac{1}{2}_{\left(  \frac{1}{2}, 0 \right)} }  & &  \frac{1}{2}_{\left( 0, \frac{1}{2} \right)}      \\
 &&&&\\
 3   &       { 0_{\left( 1,0 \right)}}\quad &  &     { {1_{(\frac{1}{2},\frac{1}{2})}}}  ,\,  0_{\left(  \frac{1}{2}, \frac{1}{2} \right)}   &  &  \qquad 0_{\left( 0,1 \right)}  \\
 &&&&\\
\frac{7}{2}  &                &        {\frac{1}{2}_{\left( 1, \frac{1}{2} \right)} }  & &  \frac{1}{2}_{\left(  \frac{1}{2},1 \right)}      \\
&&&&\\
 4  &            &      &    0_{\left( 1,1 \right)} \\
 &&&-0_{(0,0)},\,-1_{(0,0)}&\\
 &&&&\\
 \frac{9}{2}&&-\frac{1}{2}_{(\frac{1}{2},0)}&&-\frac{1}{2}_{(0,\frac{1}{2})}\\
 &&&&\\
 5 &&&-0_{(\frac{1}{2},\frac{1}{2})}&\\
 &&&&\\
 \hline
  r               &   1\qquad  & \frac{1}{2}       &   0    & -\frac{1}{2}  & \qquad-1
      \end{array}
\end{equation}
\end{itemize}
The   $\hat{\CC}_{(0,0)}$ multiplet has as its primary a ``length two'' scalar $\mathcal{T} = \bar\phi  \phi -   \MM_{{\bf {1}}}$\footnote{To obtain this precise form of the eigenvector of the Dilatation operator, an one-loop calculation is needed \cite{Gadde:2010zi}. See Section \ref{sec:elementaryexit}.}. On the other hand in generic theories $\hat{\EE}_{r}$ and $\hat{\BB}_{R}$ can have arbitrary length.
Moreover, operators that obey these shortening conditions are protected (their anomalous dimensions will be zero) and they will serve as possible vacua of the $\mathcal{N}=2$ spin chains.

\medskip

There is one more multiplet that is not 1/2 BPS, but 1/4 BPS which we wish to mention here, the $\mathcal{C}_{0,r(0,0)}$ which obeys two  $\mathcal{C}$-type conditions and has a primary with  $R=0$ and $\Delta=2 + r$.
The primaries of these multiplets are $\tr \left( \mathcal{T}\bar{\phi}^\ell \right)$ with $\ell=r$ and are also protected operators. The  $\tr \left( \mathcal{T}\phi^\ell \right)$ operators correspond to the KK tower of a 7D sugra multiplet \cite{Gadde:2009dj}.
They will describe states with a gapless magnon with momentum $p=0$, as we will see in the next section.

\medskip

As we learned in the previous section the recombination rules of the different short multiplets can teach us lessons concerning the dynamic of the CFTs. The recombination rules for ${\cal N}=2$ superconformal algebra are \cite{Dolan:2002zh}
\begin{eqnarray}
\mathcal{A}_{R,r(j,\bar{j})}^{2R+r+2j+2} & \simeq & \CC_{R,r(j,\bar{j})}\oplus\CC_{R+\frac{1}{2},r+\frac{1}{2}(j-\frac{1}{2},\bar{j})}
\label{recomb2}\\
\mathcal{A}_{R,r(j,\bar{j})}^{2R-r+2\bar j+2} & \simeq & \bar\CC_{R,r(j,\bar{j})}\oplus\bar\CC_{R+\frac{1}{2},r-\frac{1}{2}(j,\bar{j}-\frac{1}{2})}
\label{eq:3rd recomb}\\
\mathcal{A}_{R,j-\bar{j}(j,\bar{j})}^{2R+j+\bar{j}+2} & \simeq & \hat{\CC}_{R(j,\bar{j})}\oplus\hat{\CC}_{R+\frac{1}{2}(j-\frac{1}{2},\bar{j})}\oplus\hat{\CC}_{R+\frac{1}{2}(j,\bar{j}-\frac{1}{2})}\oplus\hat{\CC}_{R+1(j-\frac{1}{2},\bar{j}-\frac{1}{2})} \,.
\label{recomb1}
\end{eqnarray}
Note that the  $\EE_{r}$, and their conjugates,
$\hat{\BB}_{R}$ with $R=1/2,1,3/2$, ${\DD}_{R}$  with $R=0,1/2$ and their conjugate
multiplets can
never appear at the right hand side of a recombination rule\footnote{To see this statement one needs to know that for the special cases $\hat{\CC}_{R(j,-\frac{1}{2})}\simeq \bar{D}_{R+ \frac{1}{2}(j,0)}$, $\hat{\CC}_{R(-\frac{1}{2}, \bar{j})}\simeq {D}_{R+ \frac{1}{2}(0,\bar{j})}$ and $ {D}_{R(0,-\frac{1}{2})} \simeq \bar{D}_{R(-\frac{1}{2},0)} \simeq \hat{\BB}_{R+ \frac{1}{2}}$.} and thus are  protected. The same is true also for ${\BB}_{\frac{1}{2},r(0,\bar{j})}$ and  ${\BB}_{\frac{1}{2},r(j,0)}$.
This way we immediately learn that all Coulomb branch operators are guaranteed to be protected. Similarly all the mesonic operators (moment maps), which generate  the Higgs branch, are protected by representation theory alone.

\bigskip

The representations of $\mathcal{N}=4$ superalgebra can be decomposed to  $\mathcal{N}=2$ representations so we will not separately present them here, however, we wish to mention that  $\mathcal{N}=4$ superalgebra has one $\frac{1}{2}$-BPS multiplet
that decomposes to $\NN=2$
multiplets as \cite{Dolan:2002zh}
\begin{eqnarray}
\label{eq:N=4BPSmultipletDecomposition2N=2}
\BB_{[0,p,0]}^{\frac{1}{2},\frac{1}{2}} & \simeq & (p+1)\hat{\BB}_{\frac{1}{2}p}\oplus\EE_{p(0,0)}\oplus\bar{\EE}_{-p(0,0)}\nonumber \\
 &  & \oplus(p-1)\hat{\CC}_{\frac{1}{2}p-1(0,0)}\oplus p(\DD_{\frac{1}{2}(p-1)(0,0)}\oplus\bar{\DD}_{\frac{1}{2}(p-1)(0,0)}\nonumber \\
 &  & \oplus\bigoplus_{k=1}^{p-2}(k+1)(\BB_{\frac{1}{2}k,p-k(0,0)}\oplus\bar{\BB}_{\frac{1}{2}k,k-p(0,0)})\nonumber \\
 &  & \oplus\bigoplus_{k=0}^{p-3}(k+1)(\CC_{\frac{1}{2}k,p-k-2(0,0)}\oplus\bar{\CC}_{\frac{1}{2}k,k-p+2(0,0)})\nonumber \\
 &  & \oplus\bigoplus_{k=0}^{p-4}\bigoplus_{l=0}^{p-k-4}(k+1)\mathcal{A}_{\frac{1}{2}k,p-k-4-2l(0,0)}^{p}\label{decomposition} 
 \, .
 \end{eqnarray}
 Here the subrscript $[q,p,s]$ denotes the Dynkin labels of $SU(4)_R$ while the superscript $\frac{1}{2},\frac{1}{2}$ is there to remind us that the multiplet is $\frac{1}{2}$-BPS.
Given the fact that  \eqref{eq:N=4BPSmultipletDecomposition2N=2} contains $\EE_r$ operators which as we saw above cannot recombine,  $\BB_{[0,p,0]}^{\frac{1}{2},\frac{1}{2}}$ contains the operators that are protected.
As we will see in the next sections, operators in $\BB_{[0,p,0]}^{\frac{1}{2},\frac{1}{2}}$ are possible vacua for the spin chain of  $\mathcal{N}=4$ SYM.
Similarly, operators in multiplets on the right hand side of \eqref{eq:N=4BPSmultipletDecomposition2N=2} are also protected for the corresponding $\mathcal{N}=2$  SCFTs and provide possible vacua for the spin chains of  $\mathcal{N}=2$ SCFTs.

The  $\frac{1}{2}$-BPS multiplet $\BB_{[0,1,0]}^{\frac{1}{2},\frac{1}{2}}$ (with $p=1$) plus derivatives is the so called singleton multiplet of $\mathcal{N}=4$ SYM and contains all the single fields of the massless  $\mathcal{N}=4$ supersymmetry representation, the $\mathcal{N}=4$ vector multiplet. This representation is used to build up the spin chains of $\mathcal{N}=4$ SYM as the state space for every lattice site is  the singleton representation.
$\mathcal{N}=2$ SCA has {\bf three} distinct irreducible singleton representations:

\begin{itemize}

\item $\mathcal{V}=   \bar{\DD}_0$ the vector multiplet with shortening condition $\Delta = -r = 1$  

\item $\bar{\mathcal{V}} = \DD_{0}$  
 the conjugate vector multiplet with $\Delta = r = 1$

\item $\mathcal{H}= \hat{\BB}_{1/2}$ the hypermultiplet (real representation) with  $\Delta = 2 R = 2$.

\medskip

\end{itemize}

\section{Spin chains of $\mathcal{N}=2$ SCFTs}
\label{sec:spinChains}

We now turn to the main purpose of these notes, to construct and study spin chains of $\mathcal{N}=2$ SCFTs.
We will always present the features of the  spin chains of  $\mathcal{N}=2$ SCFTs by comparing them with the features of the spin chains of  $\mathcal{N}=4$ SYM which we will assume the reader is familiar with. More details on the  the spin chains of  $\mathcal{N}=4$ SYM can be found in this volume in the lectures notes of Marius de Leeuw \cite{deLeeuw:2019usb}, in the Special issue \cite{Bombardelli:2016rwb}, as well as in the review \cite{Beisert:2010jr}.

\subsection{The Veneziano large N limit}
 $\mathcal{N}=4$ SYM is known to be integrable in the large $N$ limit.
 As we learned in Section \ref{sec:masslessReps},
all the fields of  $\mathcal{N}=4$ SYM are in the same $\mathcal{N}=4$ vector multiplet and they are all in the adjoint representation of the color group. Thus, we can use the usual 't Hooft large $N$ limit which is simply taken by sending the number of colors $N\rightarrow \infty$ and keeping the coupling constant $\lambda = g^2_{YM} N$ fixed.
Gauge theories with quarks, and in particular  $\mathcal{N}=2$ SCFTs, have more parameters, like the number of flavors $N_f$
 which we also have to specify how to treat the large $N$ limit. There are two possible options. The first is to send $N\rightarrow \infty$ while keeping $N_f$ fixed ($N_f << N_c$  a.k.a. quenched approximation). 
For theories with $\mathcal{N}=2$ supersymmetry the beta function is one-loop exact \cite{Seiberg:1988ur}. For example for  $\mathcal{N}=2$ SQCD, 
\be
\label{eq:betaSQCD}
\beta = \frac{g^3}{16 \pi^2} (N_f - 2 N_c)
\ee
and thus, if we take the large $N$ limit keeping $N_f$ fixed we cannot obtain a CFT.  \quad
$\mathcal{N}=2$ SCFTs and in particular SCQCD admits a {\em Veneziano} expansion:
 \begin{eqnarray*}
&N = N_c  \rightarrow \infty  \quad \mbox{and} \quad N_f  \rightarrow \infty 
\quad
  \mbox{with}   \quad \frac{N_f}{N_c}=2   \quad \mbox{and} \quad \lambda = g^2_{YM} N \quad  \mbox{kept fixed}. &
  \end{eqnarray*}
In this case it is useful to use  ``generalized double line notation'' where we draw Feynman diagrams with propagators:
 \begin{figure}[h!]
\begin{centering}
\includegraphics[scale=0.4]{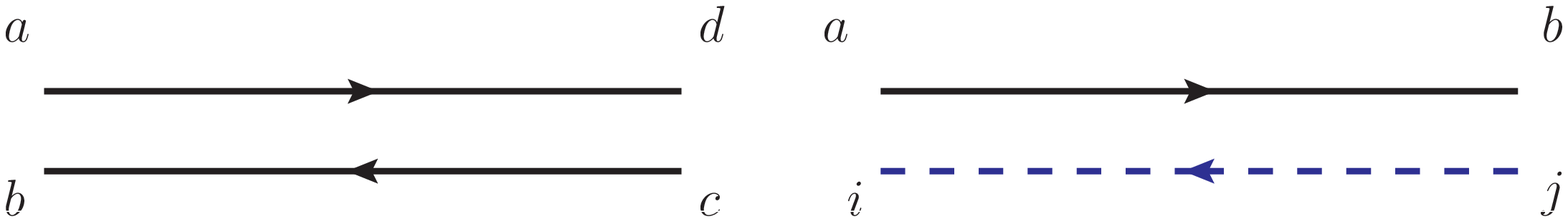}
\par\end{centering}
\end{figure}
\\
where the black line depicts the color index $a,b,c,d=1,\dots,N$ while the blue dashed line the flavor index $i,j=1,\dots,N_f$. 
Using Witten's  normalization \cite{Witten:1979kh} where we pull out a factor of $N$ in front of the single trace Lagrangian, each vertex contributes $\lambda N$, each propagator with $1 / N$ and each closed color  loop (or flavor  loop in the  Veneziano case)  contributes one more $N$.
We can quickly convince ourselves via working out a few examples that an   important feature of the the Veneziano limit, where $N_f \propto N_c$ is that
the two diagrams below are of the same order: $N^2$.  
\\
\begin{figure}[h!]
\begin{centering}
\includegraphics[scale=0.35]{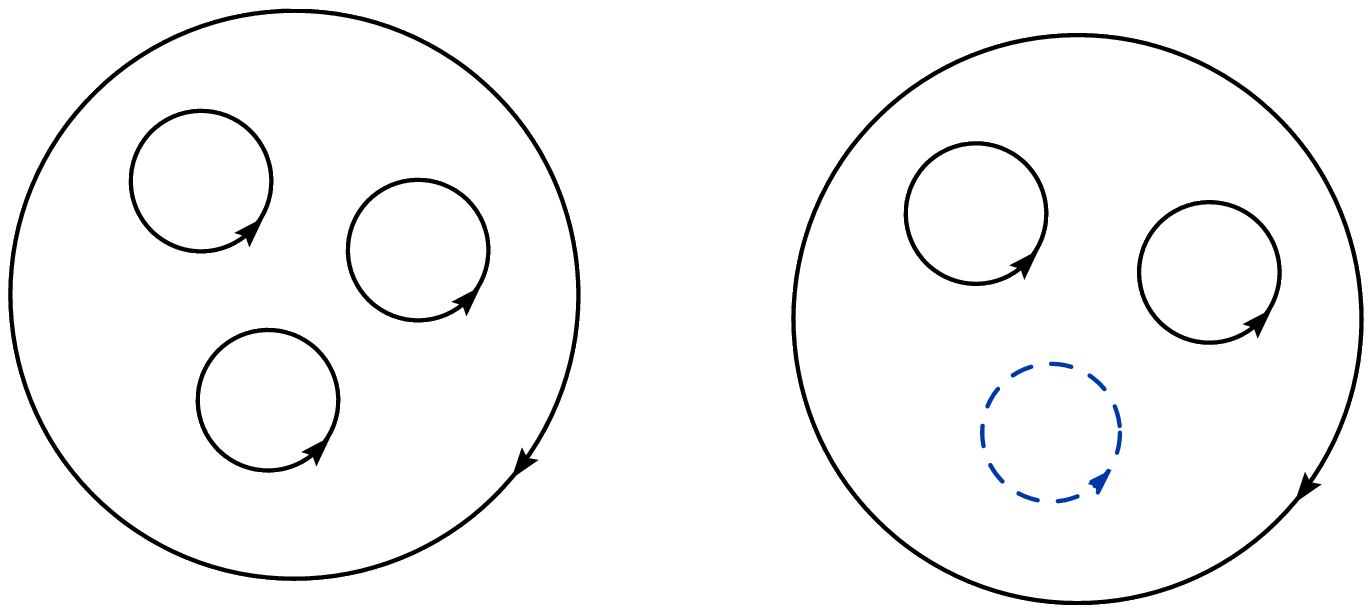}
\par\end{centering}
\end{figure} 
\\
Exactly because in the Veneziano expansion $N \propto N_f$, pure ``gluball'' type operators (with fields only in the vector multiplet) will mix with operators that contain ``mesons'' (hypers):
  \be
 \OO \sim  \Tr\left(  \phi^\ell\bar{\phi} \right) +    \Tr \left(  \phi^{\ell-1}  q_i \bar q^i  \right) \, ,
 \nonumber
  \ee
the  AdS  gravity dual will have closed string states that correspond to these   ``generalized single-trace'' operators \cite{Gadde:2009dj}
\begin{equation}
{\rm Tr}\left(\phi^{k_{1}}{\cal M}^{\ell_{1}}\phi^{k_{2}}\dots\phi^{k_{n}}{\cal M}^{\ell_{n}}\right)    
 \,,\quad   {\cal M}_{\; b}^{a}\equiv\sum_{i=1}^{N_{f}}q_{\, \, \, i}^{a}\,\bar{q}_{\, b}^{i}
 \,,\quad   a,b=1,\dots, N \, , \quad
 i=1,\dots,N_f
\nonumber \, ,
\end{equation}
where it is important to stress that when we have a quark $q$ inside the trace, we need to put  a $\bar{q}$ right after it and  flavor contract them, forming a meson in the adjoint of the color group ${\cal M}_{\; b}^{a}$, so that we pick up the leading $N$ contribution.

 \subsection{The state space}
 \label{sec:statespace}
 
 We describe the spin chains of  $\NN =4$ SYM  by allowing each  site to host a ``letter'' from the unique {\it ultrashort singleton} multiplet (= the $\frac{1}{2}$-BPS $\BB_{[0,1,0]}^{\frac{1}{2},\frac{1}{2}}$  multiplet with an arbitrary number of covariant  derivatives $\mathcal{D}_{\alpha \dot{\alpha}}$ on each
  field\footnote{Even though we omit the spinor indices and schematically write $\mathcal{D}^n$  we don't forget that antisymmetrized contractions of the covariant derivtives of the form  $\left[ \mathcal{D}_{\alpha \dot{\alpha}} \, , \, \mathcal{D}_{\beta \dot{\beta}}  \right] \propto \epsilon_{{\alpha \beta}} \bar{\mathcal{F}}_{\dot\alpha \dot\beta} +\epsilon_{\dot\alpha \dot\beta}  \mathcal{F}_{{\alpha \beta}}$ create a second field or a second cite to the spin chain, increasing it's lenght. }.): 
\[
V =  \mathcal{D}^n \left(  X, \, Y, \, \, Z,  \, \bar{X}, \,  \bar{Y}, \,  \bar{Z}, \,  \lambda^A_{\alpha},  \, \bar{\lambda}_A^{\dot\alpha}, \, \mathcal{F}_{\alpha \beta} , \, \bar{\mathcal{F}}_{\dot\alpha \dot\beta} \right)
\, .
\]
The state space of every lattice site is  $\mathcal{V}_\ell = V$ and the total space is obtained simply by the product $\otimes^L_\ell \mathcal{V}_\ell$ over all the lattice sites of the chain that are $L$.
These types of spin chains with the same  state space on every lattice site are the ones that are mostly studied  in the ``AdS/CFT integrability'' literature.

\bigskip

For $\NN =2$ SCFTs, obtaining the total space is more complicated.
The state space  at each lattice site is   spanned by 

\be
\mathcal{V}_\ell  =   \{  \mathcal{V}  \, , \quad  \bar{\mathcal{V}}     \, , { \quad  \mathcal{H}   \, , \quad   \bar{\mathcal{H} }}  \}     \nonumber
\ee
where $\mathcal{V}$ and $\mathcal{H}$ now denote the   $\DD_{0(0,0)}$  and $\hat{\BB}_{1/2}$ ultra short representations of  the $\NN =2$ SCA again with an arbitrary number of  derivatives  on each field:
\be
\nonumber
\mathcal{V} = \mathcal{D}^n \left( \phi \, , \, \lambda_{\alpha}^{\mathcal{I}} \, , \, \mathcal{F}_{\alpha \beta} \right)^a\,_b    \quad , \quad      \mathcal{H}=  \mathcal{D}^n \left(Q^{\II}  \, , \, \psi     \, , \, \bar{\tilde{\psi}}  \right)^a\,_i     \quad 
\ee
where
$a,b=1,\dots, N$ are color indices, while $i = 1,\dots, N_f$ is a flavor index.
However, 
it is very important to note that due to the large $N$ limit, the color index structure  imposes {\bf restrictions} on the total space $\otimes^L_\ell \mathcal{V}_\ell$: 
\be
  \cdots  \,  \phi   \,  \phi  \, {Q}  \, {\bar{Q}}   \,   \phi \,  \phi \,  \cdots  \nonumber
 \quad  =  \quad 
    \cdots  \,   \phi^a\,_b  \, {Q}^b\,_i  \, {\bar{Q}}^i\,_c   \,   \phi^c\,_d \,   \cdots  \nonumber
\ee
where all $a,b,c,d=1,\dots, N$ are color indices.
For example $Q\phi$ and  $\phi\bar{Q}$ are not allowed in the strict large $N$ limit. Moreover, for the SCQCD every time we have $Q$ we need to put   $\bar{Q}$ right after it so that we can flavor contract them. 
$\bar{Q}Q$ is also not allowed in the  strict  Veneziano large $N$ limit.
The field content of $\NN =2$ SQCD is summarized in  Table  \ref{table:SQCDquantumnumbers}.
We are not aware of an elegant way to describe these restrictions, other than the ``orbifolding procedure'' which will be one of the important reasons why we find it useful to think of  $\NN =2$ SCQCD as a $\check{g}\to 0$ limit of the interpolating quiver depicted in Figure \ref{interpolatingquiver}.

Note that in the case of ABJM alternating spin chains have been studied. However the ones we are dealing with now are much more complicated.

\subsection{Vacua of the spin chain}

As we learned in Section \ref{sec:SCAreps} as opposed to  $\NN=4$ SYM for which all protected operators come from the $\BB_{[0,p,0]}^{\frac{1}{2},\frac{1}{2}}$ multiplet, for  $\NN=2$  SCFTs, we have different types of the possible short $\NN=2$ SCA multiplets.
This means that we have many options for vacua.
In the case of $\mathcal{N} =4$ SYM we usually make the choice of  vacuum to be the operator $\tr Z^p$  which is a highest weight state of  $\BB_{[0,p,0]}^{\frac{1}{2},\frac{1}{2}}$ with  $\Delta = p$.
This is also known as the Berenstein-Maldacena-Nastase (BMN) vacuum (a classical string rotating in $S^5$) \cite{Berenstein:2002jq}.
An other prominent choice is the  Gubser-Klebanov-Polyakov (GKP) vacuum (classical string rotation in $AdS_5$) \cite{Gubser:2002tv}, with $\Delta -S \propto \log S$, but we will not discuss it here. It may be a good idea to also study excitations of $\NN=2$ spin chains around this vacuum.

For the $\mathcal{N}=2$ SCQCD the equivalent to the BMN vacuum is the $\tr \phi^\ell$ vacuum  with $\Delta = -  r=\ell$ ($\EE_{r}$ shortening condition) which is the only scalar operator that can have an arbitrarily long length and is protected. There is also\footnote{This can be seen either after an one-loop calculation or after the computation of the superconformal index \cite{Gadde:2009dj}.} $\tr\left(\mathcal{T} \phi^\ell\right)$ but we prefer to view $\mathcal{T}$ as an excitation in the sea of $\phi$s,  as we will see in the next section.

For the interpolating quiver depicted in Figure \ref{interpolatingquiver} there are (at least) two ``reasonable'' choices: $\hat{\BB}_R$ and $\EE_{r}$ which  one could select  as BMN-like vacuum.
The first possibility is the highest weight state of  $\hat{\BB}_R$ with $\Delta = 2R$ and corresponds to the alternating $\cdots Q \tilde{Q} Q \tilde{Q} Q \tilde{Q}  \cdots$ state. The other possibility is $\EE_{r}$ whose  highest weight state has  $\Delta = - r$.
The  $\Delta = - r$ choice of the   $\phi$-vacua  leads to two inequivalent, but degenerate vacua, one for each vector multiplet of the theory; $\tr \phi^\ell$ and $\tr \check{\phi}^\ell$.

In what comes we will concentrate to the  $\phi$-vacua. A study of the $Q$-vacuum is not yet available in the literature, but it reveals very interesting properties and  is work in progress \cite{Zoubos}.

\subsection{Elementary excitations}
\label{sec:elementaryexit}

Given the complexity of the total state space it is useful for  building intuition to begin by considering first the vacua and then states with only one elementary excitation. All multi-particle states will be constructed via scattering elementary excitations.

In the case of $\mathcal{N} =4$ SYM all excitations come from the same ($\mathcal{N} =4$ vector) multiplet. Once we make:
  \begin{itemize}
  \item the choice of  vacuum to be $\tr Z^p$ with  $\Delta - p = 0$  (= {\it magnon number}). 
\item around which there exist $8_B+8_F$ {\it elementary excitations} with  $\Delta - p = 1$:  \\
 $\lambda^A_{\alpha}$, $X,\bar{X},Y,\bar{Y}$, $\mathcal{D}_{\alpha \dot\alpha}$  with $A=1,\dots ,4$ the $SU(4)$ index,
\item all other $\Delta - p \geq 2$:  $\bar Z$, $\mathcal{F}_{\alpha \beta}$,  $\dots$ excitations correspond to {\it composite states}.
\end{itemize}
At one-loop the  dispersion relation of a single excitation  in the sea of $Z$'s is 
\be
\label{eq:sinpDispersion}
E(p)= 8 \sin^{2}\left(\frac{p}{2}\right) \, .
\ee
A derivation of \eqref{eq:sinpDispersion} was given in  the lectures  of Marius de Leeuw \cite{deLeeuw:2019usb}.
An example of a composite state for
 $\NN =4$ SYM is  $\bar Z$, which can be understood as a bound state of $X \bar X$ and $Y \bar Y$:
\bea
& \cdots  \,  Z  \,   Z   \, \textcolor{red}{ \bar Z } \,  Z  \,   Z   \,  \cdots &     \nonumber
\\
\xhookrightarrow{\quad \mathcal{H}^{(1)} \quad} & \cdots  \,  Z  \,   Z   \textcolor{red}{\left( X \bar X     +  Y \bar Y \right) } Z  \,   Z   \,  \cdots  &   \nonumber
\\
\xhookrightarrow{\quad \mathcal{H}^{(1)} \quad} & \cdots  \,  Z  \,   Z   \, \textcolor{red}{ X}  \,     Z  \,  \textcolor{red}{ \bar X}  \,     Z  \,   Z   \,  \cdots    & \nonumber
\eea
Action of the one-loop Dilatation operator (Hamiltonian) $\mathcal{H}^{(1)}$ can turn $\bar Z$ to $X \bar X$ and $Y \bar Y$ which after further actions of the one-loop Dilatation operator can separate from each other and fly apart.

\bigskip

For $\mathcal{N} =2$ SCQCD it also  happens that  all excitations come from the vector multiplet. This fact is due to the color contractions in the large $N$ limit as we discussed in the previous Subsection \ref{sec:statespace}.
  \begin{itemize}
    \item After the choice $\tr \phi^\ell$ for the  vacuum   with  $\Delta + r = 0$, 
\item   there exist  $4_B+4_F$ {\it elementary excitations} with $\Delta + r = 1$: 
\\
$\lambda^\mathcal{I}_{\alpha}$ and  $\mathcal{D}_{\alpha \dot\alpha}$  with $\mathcal{I} = 1,2$ the $SU(2)_R$ index.
\item All other  $\Delta + r  \geq 2$:  $\mathcal{M}$, $\mathcal{F}_{\alpha \beta}$ $\dots$ are {\it composite states}.
\end{itemize}
Exactly because the $\lambda^\mathcal{I}_{\alpha}$s are in the vector multiplet together with $\phi$ there are no funny restrictions on the state space. 
Moreover, the Feynman diagrams that we need to compute in order to obtain the one-loop Hamiltonian elements and in particular the ``effective vertex'' are identical\footnote{In superspace $\tr\left( W^\alpha W_\alpha\right)$ and $\tr \left( e^{-V} \bar{\Phi}e^{V}\Phi \right)$ are identical both in  $\mathcal{N} =2$  SCFTs and in  $\mathcal{N} =4$ SYM.} to the $\mathcal{N} =4$ SYM ones.
Thus, as is the case of the $\mathcal{N} =4$ SYM the  elementary excitations have energy:
\be
E_{\lambda,\mathcal{D}} (p)= 8 \sin^{2}\left( \frac{p}{2} \right) \, .
\ee
At this stage a comment is in order. The fact that there exist only   $4_B+4_F$ elementary excitations around the BMN vacuum is related to the fact that
the gravity dual of $\mathcal{N} =2$ SCQCD is a non-critical string theory  \cite{Gadde:2009dj}.

Our next step is to study of the composite states.
At this point the cautious reader may have already understood, that they have the strange property of being dimmers which occupy two sites.
For simplicity we will here discuss the one-loop scalar sector of composite dimmers.

\paragraph{One-loop scalar sector of SCQCD:}
The sub-sector with only scalar fields is closed only at one-loop.
Its neirest neighbor one-loop Hamiltonian reads \cite{Gadde:2010zi}
{\footnotesize
\be   
\mathcal{H}_{\ell,\ell+1}= \bordermatrix{
 &  \phi\phi & Q \bar{Q} &  \bar{Q}Q &  \phi Q\cr
 &&&&  \cr
\phi\phi & 2 \mathbb{I} + \mathbb{K} -2 \mathbb{P} &  \sqrt{2}  & 0 & 0 \cr
Q  \bar{Q}&  \sqrt{2}  & 2\,(2 \mathbb{I} - \mathbb{K})  & 0 & 0 \cr
\bar{Q} Q& 0 & 0 & 2 \mathbb{K} & 0 \cr
\phi Q& 0 & 0 & 0 & 2 \mathbb{I}} 
 \nonumber
 \ee}
where $\mathbb{I}$ is the identity operator, $\mathbb{P}$ the permutation operator  and $\mathbb{K}$ the trace operator and they act on the $SU(2)_R$ indices.
It is a simple exercise to diagonalise the scalar excitation   in the sea of $\phi$'s and get \cite{Gadde:2010zi}:
\be
\mathcal{T} = \bar\phi  \phi -   \MM_{{\bf {1}}}    \quad \mbox{with}   \quad  E = 4\sin^2\left(\frac{p}{2}\right)   
\ee
\be
 \widetilde{\mathcal{T} }= \bar\phi  \phi +  \MM_{{\bf {1}}}    \quad \mbox{with}   \quad  E =  8  
\ee
\be
\MM_{\bf 3}    \quad \mbox{with}   \quad   E = 8   
\ee
The singlet and the triplet under the $SU(2)_R$ mesons are defined in \eqref{M1M3}.
These states of the  one-loop  scalar sector of SCQCD should be understood as composite (dimeric - they occupy two sites) and have  $\Delta + r = 2$.
To see that $\mathcal{T}$, $ \widetilde{\mathcal{T} }$ and $\MM_{\bf 3}$ can decay to two elementary excitations $\lambda$, we need to use a higher than an one-loop element of the Hamiltonian  $ \mathcal{H}^{(L>1)}$
\bea
&  \cdots  \,   \phi \,    \phi  \textcolor{red}{ \left(\phi\bar\phi \pm \mathcal{M}_{\bf 1} \right) } \phi  \,   \phi  \,  \cdots    &   \nonumber
\\
\xhookrightarrow{\quad \mathcal{H}^{(L > 1)} \quad}   & \cdots  \,   \phi  \,  \phi   \,  \textcolor{red}{ \lambda}  \,  \textcolor{red}{ \lambda}   \,         \phi  \,    \phi   \,  \cdots   &    \nonumber
\\
\xhookrightarrow{\quad \mathcal{H}^{(1)} \quad} &  \cdots  \,   \phi  \,  \phi   \,  \textcolor{red}{ \lambda}  \,      \phi  \,  \textcolor{red}{ \lambda}  \,      \phi  \,    \phi   \,  \cdots     &\nonumber
\eea
The SU(2)$_R$ singlets
$\mathcal{T}$ and $\widetilde{\mathcal{T} }$ should be though of as bound states of  two $\lambda$s in the singlet representation of SU(2)$_R$  $\epsilon^{\alpha \beta} \lambda^{\mathcal{I}}_{\alpha}  \lambda_{{\mathcal{I}} \,\beta} \leftrightarrow \phi \left(  \bar\phi  \phi \pm  \MM_{{\bf {1}}} \right)$. 
The same can be done with $\MM_{\bf 3}$ but using two $\lambda$s in the triplet representation of SU(2)$_R$  $\epsilon^{\alpha \beta} \lambda^1_{\alpha}  \lambda^2_{\beta} \leftrightarrow \phi \mathcal{M}_{\mathbf{3}_+}$. The scattering to two-loops was studied in \cite{Gadde:2012rv}.

The reader should also note that, as claimed in the previous section, the $\tr \left( \mathcal{T}\phi^\ell \right)$ operator (corresponding to the $p=0$ case above) is also a protected operators of $\mathcal{N}=2$ SCQCD. 

\paragraph{Scalar impurities in the interpolating theory:}
 To address the complication that composite magnons in $\mathcal{N}=2$ SCQCD are dimeric,
  it is useful to  think that we are ``regularizing'' the spin chain of $\mathcal{N}=2$ SCQCD by gauging the flavor symmetry and consider the interpolating orbifold theory (SCQCD  ${\check g} \to 0$)
with $\frac{\check g}{g}$ being the regulator.
  We   regularize by inserting $ \check{\mathbf{ \phi}}$s between the ${ Q}$s giving the dimeric impurities the possibility to split:
\be
  \cdots  \,  \phi   \,  \phi  \, \textcolor{red}{Q}  \,  \textcolor{blue}{ \check{\phi}    \, \check{\phi}    \,  \cdots \,  \check{ \phi}  \,  \check{\phi}   }
     \, \textcolor{red}{\bar{Q}}   \,   \phi \,  \phi \,  \cdots  \nonumber
\ee
It is important 
 to remind the reader that in this case we have two degenerate vacua  $\tr \phi^\ell$ and  $\textcolor{blue}{\tr \check{\phi}^\ell}$.
A single excitation $Q$ in the sea of $\phi$'s interpolates between the two different vacua   
\be
\dots \, \phi \, \phi \, \phi \, \textcolor{red}{Q} \, \textcolor{blue}{ \check \phi  \,  \check \phi \,  \check \phi  \,\dots}
\ee 
and we need to necessarily consider an open spin chain, as the scalars  $\phi$ and  $\check\phi$ cannot be color contracted.
This is not a gauge invariant operator.
Nonetheless, the merit of considering this non-gauge invariant operator is that after  gauging the flavor symmetry, the $Q$s can move independently with
\be
\Delta + r =1 ( = \mbox{\it magnon number} ) \nonumber \, ,
\ee
thus, they can be interpreted as elementary magnons, bringing us
back to  $8_B+8_F$  elementary excitations with  $\Delta + r = 1$ and an AdS gravity dual that is a critical string theory.
The dispersion relation of single excitation $Q$ in the sea of $\phi$'s (which can be computed using the one-loop Hamiltonian \eqref{HamiltonianSU2})  \cite{Gadde:2010zi}:
\be
g^2 E(p)= 2(g- \check{g})^{2}+8\,g\, \check{g} \sin^{2}\left(\frac{p}{2}\right) \nonumber \,.
\ee
This  dispersion relation has a new feature compared to its $\NN= 4$ SYM counterpart, a mass gap $g^2 E(p=0) = 2(g- \check{g})^{2}$. Interestingly,
\begin{itemize}
\item at the orbifold point, where
$g= \check{g}$,  the mass gap is zero and the dispersion relation becomes identical to the $\NN= 4$ SYM one.
\item In the  SCQCD limit $\check{g}\rightarrow 0$, $E(p) = 2$ and the $Q$'s  cannot move any more and the spin chain breaks (the ${\check\phi}$s decouple),
\end{itemize}
in agreement with what we have seen above.

\subsection{Important sub-sectors}
\paragraph{Sectors with fields only in the  ${\cal N}=2$ vector multiplet:}
 As we just saw, adjoint fermions $\lambda$ in the sea  of $\phi$'s
 at one-loop  have dispersion relation and scattering matrices identical to their ${\cal N}=4$ SYM counterparts.
 Precisely the same is true for derivatives $\mathcal{D}$  in the sea  of $\phi$'s \cite{Liendo:2011xb}.
These two cases correspond to important closed sub-sectors are known as the $SU(1|1)$ and  $SU(1,1)$ sectors, respectively.
The $SU(1|1)$ sub-sector  
 is made out of one adjoint scalar and one adjoint fermion both in the $\mathcal{N}=2$ vector multiplet $( \phi ,    \lambda^{\mathcal{I}=+}_{\alpha = +})$ which are related to each other by the supercharge in the  $su(1|1)$ superalgebra.
 Similarly the  non-compact $SU(1,1)$  bosonic sector is made out of operators with 
$\phi$'s and one kind of derivative, say $\mathcal{D}_{+ \dot{+}}$. Both of them have the advantage that are closed to all-loops, simply by charge conservation and due to their symmetry and field content being identical to their ${\cal N}=4$ SYM counterparts. The reader is invited to draw the Feynman diagrams which would compute the Hamiltonians to explicitly see that, after recalling that the pieces of the action $\tr\left( W^\alpha W_\alpha\right)$ and $\tr \left( e^{-V} \bar{\Phi}e^{V}\Phi \right)$ which are used for the computation are identical both in  $\mathcal{N} =2$  SCFTs and in  $\mathcal{N} =4$ SYM.

The biggest possible sector of operators that is made  only out of fields in the ${\cal N}=2$ vector multiplet and that is closed to all-loops is the  $SU(2,1|2)$  sector \cite{Pomoni:2013poa}:
 \be
 \label{sector-fields}
\bigg \{ \phi    \, ,  \quad    \lambda^{\mathcal{I}}_{+}      \, ,    \quad   \mathcal{F}_{++}  
  \, ,    \quad 
 \mathcal{D}_{+ \dot\alpha}  
 \bigg \}\,.
\ee
Above, for simplicity we choose  $\alpha = +$  in order to get the highest-weight state of the  {\it symmetric representation} of  the $SU(2)_\alpha$ part of the Lorentz group. Clearly, all the statements that we will make below hold for  {\it any} element in  the {symmetric 
representation}. Let us also recall that $\mathcal{I}=1,2$ is the $SU(2)_R$ symmetry index.
At one-loop it is immediately clear from the Lagrangian that the scattering of two magnons is identical to scatering in $\mathcal{N}=4$ SYM. And this this sector is integrable with it's integrability working exactly like in $\mathcal{N}=4$ SYM, a statement that  remains true to any loop order \cite{Pomoni:2013poa}.

In case it is not clear to the reader, we clarify the advantage of thinking about the smallest possible and the largest possible sub-sectors. The smallest possible sub-sectors are easy to prove that they are closed and to see that they are integrable \cite{Staudacher:2004tk}, as they contain only one type of magnon. Larger sectors contain more magnons and it is harder to show integrability \cite{Beisert:2003yb}.
 However, their symmetry (their superalgebra) is large enough to completely fix their Hamiltonian at least to three-loops \cite{Beisert:2003jj,Beisert:2003ys,Zwiebel:2005er} and their all-loop scattering matrix \cite{Beisert:2005tm}.

\bigskip
 
\paragraph{Sectors with hypermultiplets:}
\paragraph{The ``$SU(2)$'' sector:}
For the interpolating quiver theory there exists a scalar, closed to all-loops sub-sector with $\Delta = 2R  - r$  and $j=\bar{j}=0$ which is made out of the color adjoints $\phi$ and  $\check{\phi}$ and the bifundamentals $Q$ and $\tilde{Q}$.
We will refer to it as the ``$SU(2)$'' sector, because although it resembles a lot the $SU(2)$ sector of $\mathcal{N}=4$, for  $\mathcal{N}=2$  gauge theories there is no $SU(2)$ symmetry that rotates the different species into one another.

\medskip

The one-loop Hamiltonian in this ``$SU(2)$'' sector is nearest neighbour type
{\scriptsize{
\begin{eqnarray}
\label{HamiltonianSU2}
 & \mathcal{H}_{\ell,\ell+1}=
 & \bordermatrix{
&\phi \phi& Q\tilde{Q} & \check{\phi} \check{\phi} &\tilde{Q}Q & \phi Q & Q  \check{\phi} &  \check{\phi}\tilde{Q} &\tilde{Q}\phi \cr
&&&& \cr
\phi \phi& 0 & 0 & 0 & 0 & 0 & 0 & 0 & 0
\cr
Q\tilde{Q} & 0  & 0 & 0 & 0 & 0 & 0 & 0 & 0\cr
\check{\phi} \check{\phi} & 0 & 0 &0&0 & 0 & 0 & 0 & 0\cr
\tilde{Q}Q & 0 & 0 & 0  &0 & 0 & 0 & 0 & 0\cr
\phi Q & 0 & 0 & 0 & 0 & 2 & -2\kappa & 0 & 0\cr
Q\check{\phi} & 0 & 0 & 0 & 0 & -2\kappa & 2\kappa^{2} & 0 & 0\cr
\check{\phi}\tilde{Q} & 0 & 0 & 0 & 0 & 0 & 0 & 2\kappa^{2} & -2\kappa\cr
\tilde{Q}\phi & 0 & 0 & 0 & 0 & 0 & 0 & -2\kappa & 2} 
\end{eqnarray}}}
with $\kappa = \frac{\check{g}}{g}$.

As discussed before, there are only two  $\Delta = - r$ operators which correspond to the two inequivalent, but degenerate $\phi$-vacua: $\tr \left( \phi^\ell\right) $ and $\tr \left( \check{\phi}^\ell\right)$ {vacua}. In these two  $\phi$-vacua we can have two inequivalent  $\Delta + r=1$ excitations $Q$ and $\tilde{Q}$  which {interpolate between the two different vacua} as
\bea
\label{Qinphi}
&\cdots \phi \,  \phi \,   \phi  \,  Q \,  \check\phi \,  \check\phi \,   \check\phi  \cdots&
\\  \nonumber
&\cdots  \check\phi \,  \check\phi \,   \check\phi   \,  \tilde{Q} \,  \phi \,  \phi \,   \phi \cdots&
\eea
and have the  dispersion relation
\be
g^2 E(p)= 2(g- \check{g})^{2}+8\,g\, \check{g} \sin^{2}\left(\frac{p}{2}\right) \, .
\ee
 At the two magnon level $\Delta + r=2$, there exist two different scattering matrices (for the two different boundary conditions), 
\bea
& S \quad  \mbox{for} \quad \cdots \phi \,  \phi \,   \phi  \,  Q \,  \check\phi \,  \check\phi \,   \check\phi  \cdots \check\phi \,  \check\phi \,   \check\phi   \,  \tilde{Q} \,  \phi \,  \phi \,   \phi \cdots&
 \nonumber \\  \nonumber
&\tilde{S} \quad  \mbox{for} \quad\cdots  \check\phi \,  \check\phi \,   \check\phi   \,  \tilde{Q} \,  \phi \,  \phi \,   \phi \cdots \phi \,  \phi \,   \phi  \,  Q \,  \check\phi \,  \check\phi \,   \check\phi \cdots&
\, ,
\eea
which can be derived as in the lectures \cite{deLeeuw:2019usb},
\be  S\left(p_1 , p_2\right)=-\frac{1+e^{i p_1+i p_2}-2 \kappa e^{i p_2}}{1+e^{i p_1+i p_2}  -2 \kappa e^{i p_1} } 
\ee
and
\be
 \tilde{S}\left(p_1 , p_2\right) = -\frac{1+e^{i p_1+i p_2}-2 \kappa^{-1}e^{i p_2}}{1 +e^{i p_1+i p_2}  -2 \kappa^{-1}e^{i p_1}}  \,.
\ee
The have the form of
the scattering matrix of the XXZ spin chain with anisotropy parameter $\Delta =\kappa$ in the first case and $\Delta =1/\kappa$ in the second.
Given the fact that $S \neq \tilde{S}$ the standard YBE is not satisfied
\be
S \, \tilde{S} \, S \neq \, \tilde{S} \, S \, \tilde{S}.
\ee
This may suggest that $\mathcal{N}=2$ SCFTs are not integrable, at least in the usual (rational) way  because already at one-loop  (as opposed to $\mathcal{N} =4$ SYM) the scattering matrix in  scalar sector \cite{Gadde:2010zi} did not obey the usual YBE.  However, the question of integrability is not so simple to answer and it should be though through more carefully.  The XXZ spin chain is a $q$-deformation of the XXX spin chain and also integrable, however, the ``$SU(2)$'' sector of $\mathcal{N}=2$ SCFTs seems to correspond to  a nontrivial (twisted) superposition of two different  XXZ spin chains, one with   $\Delta = \kappa $  and one with $\Delta =1/\kappa$.
This type of $q$-deformed structure seems to  remain in higher loops. Via explicit Feynman diagram computations, we have checked that up to 3-loops in \cite{Pomoni:2011jj}.

At this point we cannot resist but to mention the following fact, which will appear in  \cite{Zoubos}.
The results  which we have up to now presented are derived for excitations only around the  $\phi$-vacua. A study around  the $Q$-vacuum for this ``$SU(2)$'' sector, brings to light the following intriguing results. Allowing for a single $\phi$ excitation to move in the
sea of $\cdots Q \tilde{Q} Q \tilde{Q} Q \tilde{Q}  \cdots$
 and using \eqref{HamiltonianSU2} we can derive the dispersion relation
\be
E_{\phi}(p) = 2(1+\kappa^2) - 2\sqrt{(1+\kappa^2)^2 - 2\kappa^2 \sin^2 p} \, .
\ee
This looks like the dispersion relation of an elliptic system! Already with just an one-loop computation!

\paragraph{The ``SU(3$|$2)'' sub-sector:}
As first discussed in \cite{Liendo:2011xb}, the bigger sector which includes bifundamental magnons and is similar to the  SU(3$|$2) sub-sector of $\mathcal{N}=4$ SYM is
\be 
\bigg \{\f,\check{\phi},\bar{\psi}_{\dot{\alpha} \hat{\mathcal{I}}=\hat{1}},\bar{\tilde{\psi}}_{\dot{\alpha} \hat{\mathcal{I}}=\hat{1}}, Q_{\mathcal{I}\, \hat{\mathcal{I}}=\hat{1}},\bar{Q}_{\mathcal{I}\, \hat{\mathcal{I}}=\hat{1}}\bigg\}\,.
\ee
The merit of this  ``SU(3$|$2)'' sub-sector is that like for  $\mathcal{N}=4$ SYM
\cite{Beisert:2003jj,Beisert:2005tm}, it is possible to fix the Hamiltonian and the scattering matrix simply by symmetry arguments.
We will refer to it as the ``SU(3$|$2)''  sub-sector to remember that it includes 3 bosons and two fermions. There is no SU(3) symmetry, we have only an SU(2$|$2), which is importantly preserved after the choice of the  $\phi$-vacum.
\paragraph{$\bar{\tilde{\psi}}$ $\bar{\psi}$ and  $\bar{\psi}$ $\bar{\tilde{\psi}}$ scattering:}
The index structure of the fields implies that there cannot be any transmission, $\bar{\tilde{\psi}}$ must always be to the left of $\bar{\psi}$, the process is pure reflection.
The one-loop results of \cite{Liendo:2011xb} for the four different combinations of fields and indices are summarised in Table \ref{S1},
\begin{table}[h]
\begin{center}
\begin{tabular}{|l |c| c| c| c| c| c|}
 \hline
 Incoming & Sector & Scattering Matrix  \\
 \hline
 $\bar{\tilde{\psi}}\bar{\psi}$ & $1_{\dot\alpha}\otimes 3_L$  & $S(p_1,p_2,\kappa)$   \\
 $\bar{\tilde{\psi}}\bar{\psi}$ & $3_{\dot\alpha}\otimes 3_L$  & -1   \\
 $\bar{\psi}\bar{\tilde{\psi}}$ & $1_{\dot\alpha}\otimes 3_L$  & $S(p_1,p_2,1/\kappa)$   \\
 $\bar{\psi}\bar{\tilde{\psi}}$ & $3_{\dot\alpha}\otimes 3_L$  & -1   \\
\hline
\end{tabular}
\end{center}
\caption{\label{S1} \it Components of the S-matrix in the ``SU(3$|$2)''   sub-sector.}
\end{table}

\paragraph{$\bar{\psi} Q$, $Q\bar{\psi}$, $\bar{Q}\bar{\tilde{\psi}}$ and $\bar{\tilde{\psi}}\bar{Q}$ scattering:}

These processes are a little bit more interesting because we can have reflection and transmission. 
Taking into account all four combinations we obtain
\begin{table}[h]
\begin{center}
\begin{tabular}{|l |c| c| c| c| c| c|}
 \hline
 Incoming & $T$ and $R$ matrices  \\
 \hline
 $\bar{\psi} Q$        & $T(p_1,p_2,\kappa),R(p_1,p_2,\kappa)$      \\
 $Q \bar{\psi}$        & $T(p_1,p_2,1/\kappa),R(p_1,p_2,1/\kappa)$  \\
 $\bar{Q}\bar{\tilde{\psi}}$  & $T(p_1,p_2,\kappa),R(p_1,p_2,\kappa)$      \\
 $\bar{\tilde{\psi}}\bar{Q}$  & $T(p_1,p_2,1/\kappa),R(p_1,p_2,1/\kappa)$  \\
\hline
\end{tabular}
\end{center}
\caption{
\label{S2}
\it Transmission and reflection coefficients  in the ``SU(3$|$2)''  sub-sector.}
\end{table}
where
\bea 
T(p_1,p_2) & = & -\frac{1-e^{-ip_2+ip_1}}{\kappa e^{-ip_2} + \kappa e^{ip_1} -2}\, ,
\\
R(p_1,p_2) & = &  -\frac{1-\kappa e^{-ip_2} -\kappa e^{ip_1} +e^{-ip_2+ip_1}}{\kappa e^{-ip_2} + \kappa e^{ip_1} -2}\,.
\eea

\bigskip

To summarise, the take home message from this section and in particular the Tables \ref{S1} and \ref{S2}  is that:
\begin{itemize}
\item
the scattering  of $\mathcal{V} \, \mathcal{V}$  in the $\phi$-vacuum is identically the same as in  $\NN =4$ SYM,
\item
the scattering  of $\bar{\mathcal{H}} \,  \mathcal{H}$ in the $\phi$-vacuum has scattering matrix $S(p_1,p_2,\kappa)$, while  the $\mathcal{H} \, \bar{\mathcal{H}}$  scattering has  $S(p_1,p_2,1/\kappa)$.
\end{itemize}

  \subsection{All-loop dispersion relation and Scattering Matrix}
  \label{sec:All-loopS}
  
Beisert's symmetry argument \cite{Beisert:2005tm} and the derivation of the  all-loop dispersion relations and the all-loop scattering matrix  for $\NN =4$ SYM was given in the lectures of
  Olof Ohlsson Sax on  ``Factorised scattering in AdS/CFT'' in this school.
  Here we will only recall the essential elements which render  this derivation possible and emphasise the differences between $\NN =4$ SYM and the
   $\NN =2$ SCFTs in which we are interested. The fact that  the all-loop dispersion relation and scattering matrix are also possible to derive  for   $\NN =2$ SCFTs was shown in  \cite{Gadde:2010ku}, which we will follow and where the interested reader should turn for further details.
    \begin{table}[h!]
\begin{centering}
\begin{tabular}{ccc|cc}
 & $SU(2)_{\dot{\alpha}}$ & \multicolumn{1}{c}{$SU(2)_{R}$} & $SU(2)_{\alpha}$ & $SU(2)_L$\tabularnewline
\cline{2-5} 
\multicolumn{1}{c|}{$SU(2)_{\dot{\alpha}}$} & \multicolumn{1}{c|}{$\mathcal{L}_{\:\dot{\beta}}^{\dot{\alpha}}$} 
&  \multicolumn{1}{c|}{$\bar{\mathcal{Q}}_{\: \mathcal{J}}^{\dot{\alpha}}$} 
&  \multicolumn{1}{c|}{$\mathcal{P}_{\:\beta}^{\dot{\alpha}}$} 
&  \multicolumn{1}{c|}{$\bar{\mathcal{Q}}_{\: \hat{\mathcal{J}}}^{\dot{\alpha}}$}
\tabularnewline
\cline{2-5} 
\multicolumn{1}{c|}{$SU(2)_R$} & \multicolumn{1}{c|}{$\mathcal{S}_{\:\dot{\beta}}^{\mathcal{I}}$} & $\mathcal{R}_{\: \mathcal{J}}^{\mathcal{I}}$ &  \multicolumn{1}{c|}{ $  \mathcal{Q}_{\:\beta}^{\mathcal{I} }    $ } & 
 \multicolumn{1}{c|}{$\mathcal{R}_{\: \hat{\mathcal{J}}}^{\mathcal{I}}$}
\tabularnewline
\cline{2-5} 
 \multicolumn{1}{c|}{$SU(2)_{\alpha}$} 
 &  \multicolumn{1}{c|}{$\mathcal{P}_{\:\dot{\beta}}^{\alpha}$}
& $\mathcal{Q}_{\: \mathcal{J}}^{\alpha}$ & \multicolumn{1}{c|}{$\mathcal{L}_{\:\beta}^{\alpha}$} & \multicolumn{1}{c|}{$\mathcal{Q}_{\: \hat{\mathcal{J}}}^{\alpha}$}
\tabularnewline
\cline{2-5} 
 \multicolumn{1}{c|}{$SU(2)_L$} &  \multicolumn{1}{c|}{$\bar{\mathcal{Q}}_{\:\dot{\beta}}^{\hat{\mathcal{I}}}$} &  \multicolumn{1}{c|}{$\mathcal{R}_{\: \mathcal{J}}^{\hat{\mathcal{I}}}$} 
& \multicolumn{1}{c|}{$\mathcal{S}_{\:\beta}^{\hat{\mathcal{I}}}$} 
& \multicolumn{1}{c|}{$\mathcal{R}_{\: \hat{\mathcal{J}}}^{\hat{\mathcal{I}}}$}
\tabularnewline
\cline{2-5} 
\end{tabular}
\par\end{centering}
\caption{\it The generators of $PSU(2,2|4)$. }
\label{GeneratorsPSU224}
\end{table}  

$\NN =4$ SYM enjoys the full $PSU(2,2|4)$ symmetry, the generators of which are summarised in Table \ref{GeneratorsPSU224}, using a perhaps unusual $\mathcal{N}=2$ notation.
As we have already discussed in the previous section, to describe magnons we have to first choose a vacuum, around which we will construct the exited states.  We will choose the BMN vacuum $\tr Z^\ell$.
This choice of the vacuum breaks half of the symmetries as depicted in the Table \ref{table:brokenN=4}.
\be
PSU(2,2|4) \xrightarrow{\, \text{BMN vac.} \,} PSU(2|2) \times PSU(2|2) \times \mathbb{R}  \xrightarrow{\, \text{Beisert} \,} SU(2|2) \times SU(2|2) \times \mathbb{R}
\ee
where $\mathbb{R}$ corresponds to the Hamiltonian. Beisert's idea was to allow for a central extension of the algebra which he showed is enough to fix the form of the dispersion relation and the scattering matrix.
\\
\begin{table}[h!]
\begin{centering}
\begin{tabular}{ccc|cc}
 & $SU(2)_{\dot{\alpha}}$ & \multicolumn{1}{c}{$SU(2)_{R}$} & $SU(2)_{\alpha}$ & $SU(2)_L$\tabularnewline
\cline{2-3} 
\multicolumn{1}{c|}{$SU(2)_{\dot{\alpha}}$} & \multicolumn{1}{c|}{$\mathcal{L}_{\:\dot{\beta}}^{\dot{\alpha}}$} & $\mathcal{Q}_{\: \mathcal{J}}^{\dot{\alpha}}$ & $\mathcal{D}_{\:\beta}^{\dagger\dot{\alpha}}$ & $\lambda_{\: \hat{\mathcal{J}}}^{\dagger\dot{\alpha}}$\tabularnewline
\cline{2-3} 
\multicolumn{1}{c|}{$SU(2)_R$} & \multicolumn{1}{c|}{$\mathcal{S}_{\:\dot{\beta}}^{\mathcal{I}}$} & $\mathcal{R}_{\: \mathcal{J}}^{\mathcal{I}}$ & $\lambda_{\:\beta}^{\dagger \mathcal{I}}$ & $\mathcal{X}_{\: \hat{\mathcal{J}}}^{\dagger \mathcal{I}}$\tabularnewline
\cline{2-5} 
$SU(2)_{\alpha}$ & $\mathcal{D}_{\:\dot{\beta}}^{\alpha}$ & $\lambda_{\: \mathcal{J}}^{\alpha}$ & \multicolumn{1}{c|}{$\mathcal{L}_{\:\beta}^{\alpha}$} & \multicolumn{1}{c|}{$\mathcal{Q}_{\: \hat{\mathcal{J}}}^{\alpha}$}\tabularnewline
\cline{4-5} 
$SU(2)_L$ & $\lambda_{\:\dot{\beta}}^{\hat{\mathcal{I}}}$ & $\mathcal{X}_{\: \mathcal{J}}^{\hat{\mathcal{I}}}$ & \multicolumn{1}{c|}{$\mathcal{S}_{\:\beta}^{\hat{\mathcal{I}}}$} & \multicolumn{1}{c|}{$\mathcal{R}_{\: \hat{\mathcal{J}}}^{\hat{\mathcal{I}}}$}\tabularnewline
\cline{4-5} 
\end{tabular}
\par\end{centering}
\caption{\it After choosing the vacuum: broken generators become gapless  magnons.}
\label{table:brokenN=4}
\end{table}

\begin{theorem}
 The broken generators (Goldstone excitations)  correspond to ``gapless  magnons''.
 These magnons transform in the fundamental of $SU(2|2)$ and have dispersion relation
\be
E(p) = \Delta-|r|   =   \sqrt{1+ h\left( g \right)\sin^{2}\left(\frac{p}{2}\right)   }
\ee
The two-body scattering matrix is fixed by Beisert's centrally extended $SU(2|2)\times SU(2|2)$ symmetry.  
\end{theorem}
Note that  Beisert's centrally extended $SU(2|2)$ symmetry is enough to fix the dispersion relation and the scattering matrix up to a single function $h(g)$.
For  $\NN =4$ SYM, explicit Feynman diagram computations (up to 5-loops \cite{Sieg:2010jt}) and string theory computations using AdS/CFT (up to one-loop) give
\be
\label{eq:N=4h}
h\left( g \right) = g^2 \, .
\ee
At this moment we wish to stress that there is no way to show \eqref{eq:N=4h} in any way other than   explicit   Feynman diagram  computations! Strictly speaking  \eqref{eq:N=4h} is an input, an assumption, in the  $\NN =4$ SYM integrability business.

\bigskip

   \begin{table}[h!]
\begin{centering}
\begin{tabular}{ccccc}
 & $SU(2)_{\dot{\alpha}}$ & \multicolumn{1}{c}{$SU(2)_{R}$} & $SU(2)_{\alpha}$ & $SU(2)_L$\tabularnewline
\cline{2-4} 
\multicolumn{1}{c|}{$SU(2)_{\dot{\alpha}}$} & \multicolumn{1}{c|}{$\mathcal{L}_{\:\dot{\beta}}^{\dot{\alpha}}$} 
&  \multicolumn{1}{c|}{$\mathcal{Q}_{\: \mathcal{J}}^{\dot{\alpha}}$} 
&  \multicolumn{1}{c|}{$\mathcal{P}_{\:\beta}^{\dot{\alpha}}$} 
&  
\tabularnewline
\cline{2-4} 
\multicolumn{1}{c|}{$SU(2)_R$} & \multicolumn{1}{c|}{$\mathcal{S}_{\:\dot{\beta}}^{\mathcal{I}}$} & \multicolumn{1}{c|}{ $\mathcal{R}_{\: \mathcal{J}}^{\mathcal{I}}$} &  \multicolumn{1}{c|}{$\mathcal{Q}_{\:\beta}^{\mathcal{I}} $} & 
\tabularnewline
\cline{2-4} 
 \multicolumn{1}{c|}{$SU(2)_{\alpha}$} 
 &  \multicolumn{1}{c|}{$\mathcal{P}_{\:\dot{\beta}}^{\alpha}$}
&  \multicolumn{1}{c|}{$\mathcal{Q}_{\: \mathcal{J}}^{\alpha}$} & \multicolumn{1}{c|}{$\mathcal{L}_{\:\beta}^{\alpha}$} & 
\tabularnewline
\cline{2-5} 
$SU(2)_L$ & & &  \multicolumn{1}{c|}{}
& \multicolumn{1}{c|}{$\mathcal{R}_{\: \hat{\mathcal{J}}}^{\hat{\mathcal{I}}}$}
\tabularnewline
\cline{5-5} 
\end{tabular}
\par\end{centering}
\caption{\it The superconformal algebra plus global symmetry.}
\label{N=2UnBrocken}
\end{table}

For  the interpolating quiver (depicted in Figure \ref{interpolatingquiver}) we begin with the full  $\NN =2$ SCA  $SU(2,2|2)$  plus an extra    $SU(2)_L$ global symmetry  (see Appendix \ref{sec:orbifold}).
  The choice of the $\phi$-vacuum
breaks the symmetry down to $SU(2_{\alpha})\times SU(2_{\hat{I}})\times SU(2_{\dot{\alpha}}|2_{I})$ and we obtain the excitations/magnons depicted in Table \ref{N=2Brocken}.
For generic $\NN=2$ SCFTs which enjoy just the  $SU(2,2|2)$ SCA, after choosing the vacuum to be $\tr \phi^\ell$ we break the symmetry down to $SU(2_{\dot\alpha}|2_R) \times SU(2)_{\alpha}$.
Comparing this with  $\NN=4$ SYM we can say the following.

\begin{table}[h]
\begin{centering}
\begin{tabular}{ccc|cc}
 & $SU(2)_{\dot{\alpha}}$ & \multicolumn{1}{c}{$SU(2)_{R}$} & $SU(2)_{\alpha}$ & $SU(2)_L$\tabularnewline
\cline{2-3} 
\multicolumn{1}{c|}{$SU(2)_{\dot{\alpha}}$} & \multicolumn{1}{c|}{$\mathcal{L}_{\:\dot{\beta}}^{\dot{\alpha}}$} & $\mathcal{Q}_{\: \mathcal{J}}^{\dot{\alpha}}$ & $\mathcal{\mathcal{D}}_{\:\beta}^{\dagger\dot{\alpha}}$ & ${  \psi_{\: \hat{\mathcal{J}}}^{\dagger\dot{\alpha}}   }$\tabularnewline
\cline{2-3} 
\multicolumn{1}{c|}{$SU(2)_R$} & \multicolumn{1}{c|}{$\mathcal{S}_{\:\dot{\beta}}^{\mathcal{I}}$} & $\mathcal{R}_{\: \mathcal{J}}^{\mathcal{I}}$ & $\lambda_{\:\beta}^{\dagger \mathcal{I}}$ & ${    \bar{Q}_{\: \hat{\mathcal{J}}}^{\, \mathcal{I}}   }$\tabularnewline
\cline{2-5} 
$SU(2)_{\alpha}$ & $\mathcal{D}_{\:\dot{\beta}}^{\alpha}$ & $\lambda_{\: \mathcal{J}}^{\alpha}$ & \multicolumn{1}{c|}{$\mathcal{L}_{\:\beta}^{\alpha}$} & \multicolumn{1}{c|}{
}\tabularnewline
\cline{4-5} 
$SU(2)_L$ & ${   \psi_{\:\dot{\beta}}^{\hat{\mathcal{I}}}   }$ & ${ Q_{\: \mathcal{J}}^{\hat{\mathcal{I}}}  }$ 
& \multicolumn{1}{c|}{
} 
& \multicolumn{1}{c|}{$\mathcal{R}_{\: \hat{\mathcal{J}}}^{\hat{\mathcal{I}}}$}\tabularnewline
\cline{4-5} 
\end{tabular}
\par\end{centering}
\caption{\it After choosing the vacuum: broken generators become gapless  magnons and we have some extra magnons that do not come from broken generators.}
\label{N=2Brocken}
\end{table}

The broken generators, as is the case for $\NN =4$ SYM, correspond to  Goldstone excitations and lead to  gapless magnons. These magnons come from the $\mathcal{N}=2$ vector multiplet and have the same dispersion relation as the excitations of $\NN =4$ SYM,
\be
E_{\lambda , \mathcal{D}} (p) =   \sqrt{1+8 {\bf g}^{2}\sin^{2}\left(\frac{p}{2}\right)  } \,.
\ee
Notice that as for the $\NN =4$ SYM, the centrally extended $SU(2|2)$ symmetry is enough to fix the dispersion relation and the scattering matrix up to a single function 
${\bf g}^2 = f(g,\check{g}) = g^2 +\cdots$   in the case that we are expanding  around the  $\cdots \phi Q \check{\phi}\cdots$ or
${\bf \check{g}}^2 = \check{f}(g,\check{g}) = \check{g}^2 +\cdots$   in the case that we are expanding around the  $\cdots  \check{\phi} \bar{Q} {\phi}\cdots$.
These functions are computed up to three-loops in \cite{Mitev:2014yba}
\be
f(g, \check{g}) =  g^2 + 2\left(\check{g}^2-g^2\right) \left[6\zeta(3) g^4 + 20\zeta(5) g^4\left(\check{g}^2+3 g^2\right) + \cdots  \right]\ .  
\ee

{Non-existing generators} to start with correpsond to {non-Goldstone excitations} and thus lead to  gapped magnons with dispersion relation
\be
E_{Q , \psi}(p) =   \sqrt{1+2({\bf g}-\check{{\bf g}})^2+8{\bf g}\check{{\bf g}}\sin^{2}\left(\frac{p}{2}\right) }
\ee
where again we need to fix two functions $({\bf g}-\check{{\bf g}})^2= f_1(g,\check{g}) = ({g}-\check{{g}})^2 +\cdots$ 
and  ${\bf g}\check{{\bf g}}= f_2(g,\check{g}) = {g}\check{{g}} +\cdots$  via Feynman diagram computations. This is explicitly done up to 3-loops in \cite{Pomoni:2011jj}.

For the scattering matrix here we will only say  that it can also be computed for any $\mathcal N=2$ SCFT. 
After making the choice of {vacuum} to be $\tr \phi^\ell$,
the scattering matrix of highest weight states in $SU(2)_{\alpha}$ and $SU(2)_{L}$ is fixed  by the centrally extended $SU(2|2)$.    
This was done in \cite{Gadde:2010ku}. The scattering matrix  is also completely fixed up to two functions $f_1(g,\check{g})$ and  $ f_2(g,\check{g})$ (as above).

\medskip

We conclude this section emphasizing that, as we discussed for the one-loop approximation,
there exist two different scattering matrices (for the two different boundary conditions)
\bea
& S \quad  \mbox{for} \quad \cdots \phi \,  \phi \,   \phi  \,  Q \,  \check\phi \,  \check\phi \,   \check\phi  \cdots \check\phi \,  \check\phi \,   \check\phi   \,  \tilde{Q} \,  \phi \,  \phi \,   \phi \cdots&
 \nonumber \\  \nonumber
&\tilde{S} \quad  \mbox{for} \quad\cdots  \check\phi \,  \check\phi \,   \check\phi   \,  \tilde{Q} \,  \phi \,  \phi \,   \phi \cdots \phi \,  \phi \,   \phi  \,  Q \,  \check\phi \,  \check\phi \,   \check\phi \cdots&
\eea
with
\be  
\label{eq:SU(2|2)StildeS}
S=S \left(p_1 , p_2, \kappa\right) \qquad \mbox{and} \qquad \tilde{S}=  {S}\left(p_1 , p_2, \kappa^{-1}\right)  \, .
\ee
The matrices in \eqref{eq:SU(2|2)StildeS}, given the fact that $S \neq \tilde{S}$ do not satisfy the standard YBE 
\be
S \, \tilde{S} \, S \neq \, \tilde{S} \, S \, \tilde{S} \, ,
\ee
precisely as we found in the simplest one-loop, ``SU(2)'' sector.

\medskip

We finish this section with yet an other intriguing observation. The $SU(2|2)$ scattering matrix of  \cite{Gadde:2010ku} for $Q$ excitations in the sea of $\phi \,   \phi  \,  Q \,  \check\phi \,  \check\phi \,   \check\phi$ has precisely the same form as the
quantum double $SU(2|2)$ scattering matrix of
\cite{Beisert:2016qei}, for a certain choice of the quantum deformation parameters. Similarly,  $\tilde{Q}$ excitations in the sea of $\check\phi \,   \check\phi   \,  \tilde{Q} \,  \phi \,  \phi$ have a scattering matrix which is equal to the quantum double $SU(2|2)$ scattering matrix of
\cite{Beisert:2016qei}, for a slightly different   choice ($\kappa \leftrightarrow \kappa^{-1}$)  of the quantum deformation parameters.
This is very reminiscent of the work of \cite{Felder:1994be}  on elliptic quantum groups and  a modified, dynamical  Yang-Baxter equation.
In fact it looks very possible that this  ``SU(3$|$2)'' sub-sector is governed by a certain elliptic integrable model which is precisely based on an elliptic quantum $SU(2|2)$ and will obey only a dynamical  Yang-Baxter equation \`a la \cite{Felder:1994be}.

 \section{Where we stand}

\subsection{There is an integrable sub-sector (only vector multiplet)}
\begin{figure}[h]
   \centering
   \includegraphics[scale=0.5]{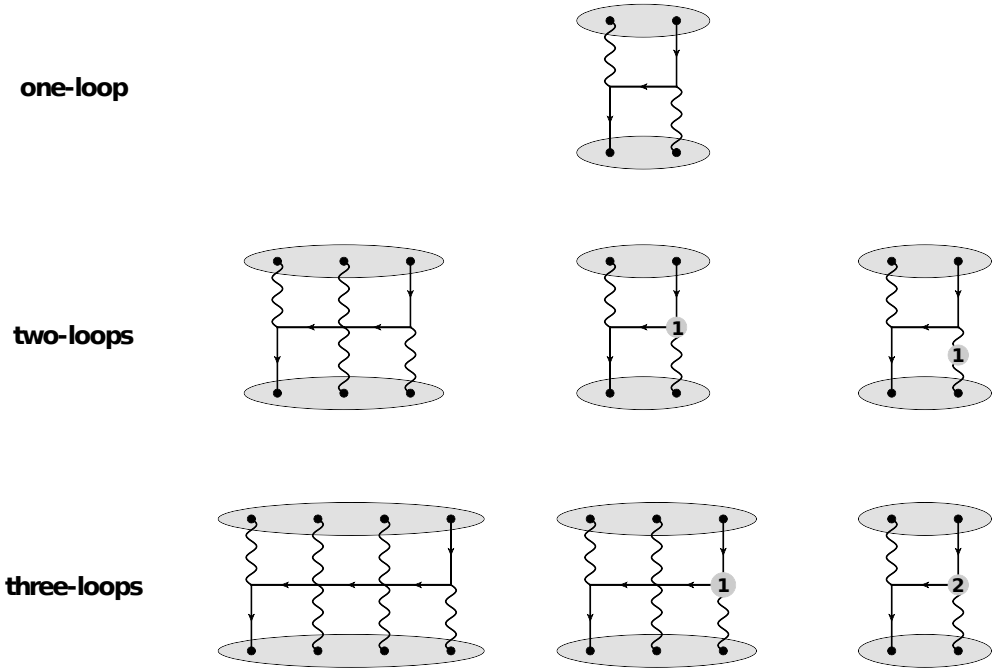} 
   \caption{\it Representatives of the Feynman diagrams we need to compute to three-loops for the Hamiltonian of the  $SU(2,1|2)$  sector. The solid lines represents the adjoint $\Phi$ chiral superfield in the $\mathcal{N}=2$ vector multiplet, while the curly lines  the real vector   superfield $V$.}
   \label{fig:1-3-loop}
\end{figure}
 We already saw that at the one-loop level if we stick to a sub-sector with only fields in one of the vector multiplets, the Hamiltonian is identical to  $\mathcal{N}=4$ SYM, $\mathcal{H}^{(1)}_{\mathcal{N}=2} = \mathcal{H}^{(1)}_{\mathcal{N}=4}$. To two-loops (second line in Figure \ref{fig:1-3-loop}) the diagram on the left is also identical to its $\mathcal{N}=4$ SYM counterpart as we only use vertices from the vector multiplet part of the Lagrangian that is identical to its counterpart in  $\mathcal{N}=4$ SYM. Leg and vertex corrections can in principle be different by finite corrections, however at one-loop they happen to be identical to $\mathcal{N}=4$ SYM, after an explicit computation \cite{Pomoni:2011jj}. Thus, to two-loops  $\mathcal{H}^{(2)}_{\mathcal{N}=2} = \mathcal{H}^{(2)}_{\mathcal{N}=4}$.
 Finally, to three-loops  (third line in Figure \ref{fig:1-3-loop}) the only diagrams\footnote{There is one diagram to three-loops which could spoil this argument, however, after more careful examination and explicit computation, this diagram does not contribute to the Hamiltonian because it is finite, as the non-renormalisation theorem of \cite{Fiamberti:2008sh,Sieg:2010tz} suggests. } that can be different from their $\mathcal{N}=4$ SYM counterparts, are of the form of the diagram depicted on the right of third line in Figure \ref{fig:1-3-loop}, which contain a two-loop leg or vertex correction
 and
  which is proportional to the one-loop Hamiltonian,
\be
\mathcal{H}^{(3)}_{\mathcal{N}=2} - \mathcal{H}^{(3)}_{\mathcal{N}=4} \propto \lambda^3 \mathcal{H}^{(1)}_{\mathcal{N}=4} \quad \Rightarrow \quad \mathcal{H}_{\mathcal{N}=2} = \mathcal{H}_{\mathcal{N}=4}(f(\lambda)) + \mathcal{O}(\lambda^4)
\,.
\nonumber
\ee
This logic can be iterated until we reach the following conclusion \cite{Pomoni:2013poa}:
\\
Every $\mathcal{N}=2$ SCFT has a  purely gluonic 
  $SU(2,1|2)$  sector,   integrable in the planar limit 
 with its integrability immediately inherited from   $\mathcal{N}=4$ SYM, as its Dilatation operator
 \be
 \nonumber
\mathcal{H}_{\mathcal{N}=2} \left( g \right) = \mathcal{H}_{\mathcal{N}=4} \left( \mathbf{g} \right)  \, .
 \ee 
The {\bf effective coupling}:
  \be
 \nonumber
\mathbf{g}^2 = f(g^2) = g^2+ g^2 \left( Z_{\mathcal{N}=2} -  Z_{\mathcal{N}=4}\right)
 \ee 
 encodes the {\bf relative  finite renormalization} of $g$ and we can either compute it via Feynman diagrams (to some loop order), or
we can compute it via Localization by comparing the 1/2-BPS circular Wilson loop,  \cite{Mitev:2014yba}:
   \be
 \nonumber
W_{\mathcal{N}=2}\left( {g}^2\right)  = W_{\mathcal{N}=4}\left(  \mathbf{g}^2 \right)  \, ,
 \ee 
 to any order we like.
 What is more, there is a longer list of observables for which this coupling substitution  trick seems to work \cite{Fiol:2015spa,Mitev:2015oty,Grozin:2015kna,Gomez:2018usu,Billo:2019fbi,Bianchi:2019dlw}!

 Using
 AdS/CFT we should understand this function $f(g^2)$  as the {\bf effective string tension}  $f(g^2) = T^2_{eff} = \left(  \frac{R^4}{(2\pi \alpha')^2} \right)_{eff}
$.
From the gravity dual side we can also check that  the coupling substitution rule works to leading order in the strong coupling limit \cite{Mitev:2015oty}.
All in all, AdS/CFT seems to suggests that all that happens in comparison to   $\mathcal{N}=4$ SYM is that   $f(g^2) = T^2_{eff}$ renormalizes! Thus, it is not too optimistic to hope is that we should be able to obtain any observable which classically resides in the factor $AdS_5 \times S^1$ of the geometry by replacing $g^2 \rightarrow f(g^2)$.

It would be very important to have more diagrammatic checks of the diagrammatic argument in \cite{Pomoni:2013poa} for the purely gluonic  SU(2,1$|$2) sector, as well as for the coupling substitution rule  \cite{Mitev:2014yba} (beyond four-loops).
What is more, it seems that this coupling substitution rule will also apply to a purely gluonic:
\begin{itemize}
  \item
   $SU(2,1|1)$ sub-sector in any $\mathcal{N}=1$ superconformal gauge theories \cite{Carstensen} and
     \item
   $SU(2,1)$ sub-sector in any $\mathcal{N}=0$ superconformal gauge theories 
 \end{itemize}
which would be worth exploring both with explicit Feynman diagram computations, as well as with symmetry arguments ( \`a la Beisert).
For $\NN=1$ SCFTs in class $\mathcal{S}_k$ most of the results that we have for   $\NN=2$ SCFTs immediately go though.
See \cite{Sadri:2005gi} for a first attempt with very interesting applications.

\subsection{Orbifolds of  $\mathcal{N}=4$ SYM are integrable (with hypers)}
In this section we will present  a very short review  of the work of \cite{Beisert:2005he}. Beisert and Roiban where able to show that the
orbifold daughters of  $\mathcal{N}=4$ SYM
 we introduced in Section \ref{subsec:Orbifolds}
are integrable. Integrability is also there  for more general orbifolds \cite{Solovyov:2007pw}, at the orbifold point, as well as for orbifolds that preserve
 $\mathcal{N}=1$ supersymmetry\footnote{Supersymmetry breaking orbifolds suffer from a Tachyon instability \cite{Dymarsky:2005uh,Dymarsky:2005nc,Pomoni:2008de}.}.
Discovering integrability for orbifold daughters of  $\mathcal{N}=4$ SYM is done through  the following steps:
\begin{itemize}
\item 
First we need to define a twist operator which commutes with the fields in the   $\mathcal{N}=2$ vector multiplet, but produces a phase for the ``twisted'' fields in the hypermultiplet  (recall equation \eqref{eq:orbifoldCondition}).

\item 
In the process of computing the anomalous dimensions of untwisted operators, no twist operator is involved and  the orbifold Hamiltonian is the same as in $\mathcal{N}=4$, as should be clear from the discussion in  in Section \ref{subsec:Orbifolds}.  Thus, the asymptotic Bethe ansatz equations for untwisted operators are the same as for $\mathcal{N}=4$ SYM.

\smallskip

\item When computing the anomalous dimensions of twisted operators, a  twist operator $\gamma$ is present but can and should be shifted  using the commutation relation \eqref{eq:orbifoldCondition} as
\be
X \, \gamma = e^{2\pi i s_{X} /2} \gamma X \, .
\ee
This equation determines the phase shift  for exchanging  $\gamma$ with $X$, which $s_{X}$ given by the $R$-symmetry of the field $X$, as in equation \eqref{eq:orbifoldCondition}.

\smallskip

\item 
We now should think of $\gamma$ as one more excitation which does not have a spectral parameter\footnote{In this sence $\gamma$ is more like $Z$ which also does not have a spectral parameter associated to it}.  Then, the scattering matrix of it with any other excitation is
\be
S_{X,\gamma} = \frac{1}{S_{\gamma, X} } =   e^{2\pi i s_{X} /2}
\ee

\item
The  Bethe equations are just a product of all the scattering matrices (equations (3.8) in  \cite{Beisert:2005he}) and schematically look like:
\be
\label{eq:OrbifBetheEqns}
\prod^{all \, exitations}  \prod^{M} S =1
\ee
where the product $ \prod^{M}$ runs over the number of magnons $M$ which our operator contains and  the product $\prod^{all \, exitations}$ over all possible types including $\gamma$. 
\end{itemize}

\subsection{Marginal deformations away from the orbifold point}

Summarising  what we have seen in the sections before,
 there are some sectors (with only fields in the vector multiplets) that are integrable, and some sectors which at least naively are not, because the  the standard  YBE is not satisfied. These are sectors that include hypermultiplets and thus will include twisted operators.

Up to now nobody has tried to  write down something like \eqref{eq:OrbifBetheEqns}
and to see if they would produce the correct anomalous dimensions, even at one-loop. It is very possible that if we can define 
a  twist operator which depends on the coupling constants (the marginal deformations away from the orbifold point) of the form
\be
X \, \gamma = \kappa^{\pm}  \,e^{2\pi i s_{X} /2} \gamma X
\ee
we could succeed in having an equation of the form of \eqref{eq:OrbifBetheEqns}.
This idea brings to mind and should be combined with the work of
\cite{Mansson:2008xv,Dlamini:2016aaa,Dlamini:2019zuk}.

\medskip

It is important to stress that even for $\mathcal{N}=4$ SYM the all-loop Beisert (spin-chain) $SU(2|2)$ S-matrix does not always satisfy the standard YBE \cite{Arutyunov:2006yd}.
Although for the world-sheet $SU(2|2)$ S-matrix  the usual YBE is satisfied, showing  integrability for the spin-chain of $\mathcal{N}=4$ requires the use of the twisted  Zamolodchikov-Faddeev (ZF) algebra and only the twisted YBE is satisfied \cite{Arutyunov:2006yd}. The reason for that   is that  single magnon states can not simply be tensored to give two-magnon states and that the transformation of the ZF basis involves the momentum operator. A similar, but more complicated approach seems to be needed for the spin chains of $\mathcal{N}=2$ SCFTs.

\subsection{Fixing the dilatation operator just with symmetry arguments}
Even though there was no time in these lectures to cover this very interesting direction,
we wish to just let the reader know that
the  $\mathcal{N}=2$ and  $\mathcal{N}=1$ SCAs are powerful enough to completely fix the ``complete one-loop Hamiltonians''.
From \cite{Liendo:2011xb,Liendo:2011wc} we know how to get the complete one-loop hamiltonians for   $\mathcal{N}=2$ and  $\mathcal{N}=1$ SCFTs purely using representation theory. Amazingly, even  $\mathcal{N}=0$ Hamiltonians can be fixed only using the conformal representation theory plus some minimal dynamical input (the multiplet recombination  in \eqref{eq:RecombinationRules4}) \cite{Liendo:2017wsn}!

It is also possible to  obtain higher loops Hamiltonians simply via using the superconformal algebra \cite{Gadde:2012rv}, a direction which needs to be pushed further.

\section{Conclusions}

In these lectures, after a broad introduction to $\mathcal{N}=2$ SCFTs, we reviewed the state of the art for spin chains the spectral problem of which computes anomalous dimensions of local operators in $\mathcal{N}=2$ SCFTs. 
We saw that there exist purely gluonic closed sub-sectors of operators (which are made out only of fields in one vector multiplet) which seem to be integrable to all-loops with their integrability immediately inherited from $\mathcal{N}=4$ SYM.  
Anomalous dimensions are obtained simply by replacing the coupling constant of $\mathcal{N}=4$ SYM by $g^2 \rightarrow f(g^2)$ the effective coupling which can be fixed from Localization.
On the other hand,  sub-sectors of operators in which hypermultiplet fields scatter seem to be $q$-deformed versions of their  $\mathcal{N}=4$ SYM counterparts, which  do not obey the usual YBE rational (or trigonometric) integrable models do.

 One of the important points we wish to stress is that even though   $\mathcal{N}=2$ spin chains spin chains look more complicated than the  $\mathcal{N}=4$ SYM ones, and maybe naively not integrable, we should not be discouraged away from their study.
 We have stressed that  it is a very good idea to think of many $\mathcal{N}=2$  SCFTs as orbifold daughters of  $\mathcal{N}=4$ SYM. At the orbifold point they are known to be integrable  \cite{Beisert:2005he} and our main task that remains is to see if we can find a more general integrable model which could incorporate both the orbifold twist plus  marginal deformations.

There is a lot of room to look for integrable models that will possibly describe such twisted sectors with hypermultiplets, away from the orbifold point.
Integrable models are understood to belong in three big classes: rational, trigonometric and elliptic. The XXX, XXZ and XYZ spin chains are examples in each class. The same way quantum integrability is based on the Yangian symmetry  $Y(\mathfrak{g})$ for rational models, quantum affine algebra $U_q^{\text{aff}}({\mathfrak{g}})$
and quantum elliptic group  $U_{q,t}^{\text{ell}}({\mathfrak{g}})$ respectively \cite{Drinfeld,Felder95ellipticquantum,Felder_1996} give the tower of conserved charges for the  trigonometric and elliptic models, respectively.
We believe that it should be possible to take this path
 via combining the insights from the works of
\cite{Gadde:2010ku,Beisert:2016qei,Felder:1994be,Arutyunov:2006yd,Zoubos}.

\bigskip

\section*{Acknowledgements}

I am grateful to Jan Peter Carstensen, Ioana Coman, Sofia Liguori,  
Alessandro Pini and Matteo Sacchi for very useful feedback on different stages of this draft.
My work is supported by the German Research Foundation (DFG) via the Emmy Noether
program ``Exact results in Gauge theories''.

\appendix

\section{Field content and symmetries of $\NN =2$}

\subsection{${\cal N} =2$ SCQCD} 

Our first example of an ${\cal N}=2$ SCFT is  ${\cal N} =2$ SQCD with gauge group $SU(N_{c})$ and $N_{f} = 2 N_c$ fundamental hypermultiplets. 
We refer to this theory as ${\cal N} = 2$ SCQCD to stress the fact that it is conformal \eqref{eq:betaSQCD}. 
Its global symmetry group is $U(N_f) \times SU(2)_R \times U(1)_r$,
where  $SU(2)_R \times U(1)_r$  is the R-symmetry subgroup of the superconformal group. 
We use indices $\II, \JJ =\pm$  for $SU(2)_R$, $i, j=1, \dots N_f$ for the flavor group $U(N_f)$
and $a,b=1, \dots N_c$ for the color group $SU(N_c)$.

\begin{table}
\begin{centering}
\begin{tabular}{|c||c|c|c|c|}
\hline 
 & $SU(N_{c})$  & $U(N_{f})$  & $SU(2)_{R}$  & $U(1)_{r}$\tabularnewline
\hline
\hline 
$\mathcal{Q}_{\alpha}^{\II}$  & \textbf{$\mathbf{1}$}  & \textbf{$\mathbf{1}$}  & \textbf{$\mathbf{{2}}$}  & $+1/2$\tabularnewline
\hline 
$\mathcal{S}_{\II \,\alpha}$ & \textbf{$\mathbf{1}$}  & \textbf{$\mathbf{1}$}  & \textbf{$\mathbf{2}$}  & $-1/2$\tabularnewline
\hline
\hline 
$A_{\mu}$  & Adj  & \textbf{$\mathbf{1}$}  & \textbf{$\mathbf{1}$}  & $0$\tabularnewline
\hline 
$\f$  & Adj  & \textbf{$\mathbf{1}$}  & \textbf{$\mathbf{1}$}  & $-1$\tabularnewline
\hline 
$\la_{\alpha}^{\II}$  & Adj  & \textbf{$\mathbf{1}$}  & \textbf{$\mathbf{2}$}  & $-1/2$\tabularnewline
\hline 
$Q_{\II}$  & $\Box$  & $\Box$  & \textbf{$\mathbf{2}$}  & $0$\tabularnewline
\hline 
$\psi_{\alpha}$  & $\Box$  & $\Box$  & \textbf{$\mathbf{1}$}  & $+1/2$\tabularnewline
\hline 
$\tilde{\psi}_{\alpha}$  & $\overline{\Box}$  & $\overline{\Box}$  & \textbf{$\mathbf{1}$}  & $+1/2$\tabularnewline
\hline
\hline 
$\MM_{\bf 1}$ & Adj + {\bf1}  & \textbf{$\mathbf{1}$}  & \textbf{$\mathbf{1}$}  & $0$\tabularnewline
\hline 
$\MM_{\bf 3}$ & Adj  + {\bf 1}& \textbf{$\mathbf{1}$}  & \textbf{$\mathbf{3}$}  & $0$\tabularnewline
\hline
\end{tabular}
\par\end{centering}
\caption{\it The field content and quantum numbers   of  $\NN=2$ SCQCD.}
\label{table:SQCDquantumnumbers} 
\end{table}

Table \ref{table:SQCDquantumnumbers}  summarizes the field content and quantum numbers  of the model. In the conventions we are using
the supercharges  ${\cal Q}^{\II}_\alpha$, $\bar {\cal Q}_{\II \, \dot \alpha}$ and their conformal
counterparts ${\cal S}_{\II \, \alpha}$, $\bar {\cal S}^{\II}_ {\dot \alpha}$
 are $SU(2)_R$ doublets with charges $\pm 1/2$ under $U(1)_r$.

Finally, we find useful  to define the flavor contracted mesonic operators 
\[
\MM_{\JJ \, \, \, b}^{\, \, \II a} \equiv \frac{1}{\sqrt{2}} Q_{\JJ \mbox{ }i}^{\mbox{ }a}\,\bar{Q}_{\mbox{ }b}^{\II \mbox{ }i} \, ,
\]  
which can be  decomposed  into the $SU(2)_{R}$ singlet and triplet combinations
  \begin{equation} \label{M1M3}
\MM_{{\bf {1}}} \equiv \MM^{\, \,  \II}_{\II}\quad\mbox{and}\quad\MM_{ {\bf {3} } \JJ  }^{\quad \II}  \equiv
\MM^{\, \, \II}_{\JJ}-\frac{1}{2}\MM^{\, \, \KK}_{\KK}\, \delta^{\II}_{\JJ} \, .
\end{equation}

\subsection{\label{sec:orbifold} $\mathbb{Z}_2$ orbifold of ${\cal N} = 4$ and the interpolating quiver}

The second main example of an ${\cal N}=2$ SCFT use discuss in these notes is the marginally deformed orbifold daughter of   ${\cal N}=4$ SYM with its quiver depicted  in Figure \ref{interpolatingquiver}.
Its field content consists of two $\NN =2$ vector multiplets
 $(\phi, \lambda_\II, A_m)$ and  $(\check \phi, \check \lambda_\II,\check A_m)$,
 and  two bifundamental hypermultiplets,  $(Q_{\II, \hat +}, \psi_{\hat +}, \tilde \psi_{\hat +})$
 and  $(Q_{\II, \hat -}, \psi_{\hat -}, \tilde \psi_{\hat -})$. Table \ref{orbifoldcharges} summarizes the field
 content and quantum numbers of the orbifold theory.
\begin{table}
\begin{centering}
\begin{tabular}{|c||c|c|c|c|c|}
\hline 
 & $SU(N_{c})_{1}$  & $SU(N_{c})_{2}$  & $SU(2)_{R}$  & $SU(2)_{L}$  & $U(1)_{R}$\tabularnewline
\hline
\hline 
$\QQ_{\alpha}^{\II}$ & \textbf{${\bf 1}$}  & \textbf{${\bf 1}$}  & \textbf{${\bf 2}$}  & \textbf{${\bf 1}$}  & +1/2\tabularnewline
\hline 
$\mathcal{S}_{\II \, \alpha}$ & \textbf{${\bf 1}$}  & \textbf{${\bf 1}$}  & \textbf{${\bf 2}$}  & \textbf{${\bf 1}$}  & --1/2\tabularnewline
\hline
\hline 
$A_{\mu}$  & Adj  & \textbf{${\bf 1}$}  & \textbf{${\bf 1}$}  & \textbf{${\bf 1}$}  & 0\tabularnewline
\hline 
$ A_{\mu}$  & \textbf{${\bf 1}$}  & Adj  & \textbf{${\bf 1}$}  & \textbf{${\bf 1}$}  & 0\tabularnewline
\hline 
$\f$  & Adj  & \textbf{${\bf 1}$}  & \textbf{${\bf 1}$}  & \textbf{${\bf 1}$}  & --1\tabularnewline
\hline 
${\f}$  & \textbf{${\bf 1}$}  & Adj  & \textbf{${\bf 1}$}  & \textbf{${\bf 1}$}  & --1\tabularnewline
\hline 
$\la^{\II}$  & Adj  & \textbf{${\bf 1}$}  & \textbf{${\bf 2}$}  & \textbf{${\bf 1}$}  & --1/2\tabularnewline
\hline 
${\la}^{\II}$  & \textbf{${\bf 1}$}  & Adj  & \textbf{${\bf 2}$}  & \textbf{${\bf 1}$}  & --1/2\tabularnewline
\hline 
$Q_{\II\hat{\II}}$  & $\Box$  & $\overline{\Box}$  & \textbf{${\bf 2}$}  & \textbf{${\bf 2}$}  & 0\tabularnewline
\hline 
$\psi_{\hat{\II}}$  & $\Box$  & $\overline{\Box}$  & \textbf{${\bf 1}$}  & \textbf{${\bf 2}$}  & +1/2\tabularnewline
\hline 
$\tilde{\psi}_{\hat{\II}}$  & $\overline{\Box}$  & $\Box$  & \textbf{${\bf 1}$}  & \textbf{${\bf 2}$}  & +1/2\tabularnewline
\hline
\end{tabular}
\par\end{centering}
\caption{\it The field content and quantum numbers   of the $\NN=2$ interpolating quiver.}
\label{orbifoldcharges}
\end{table}
The two gauge-couplings $g_{YM}$ and $\check g_{YM}$ can be independently varied
while preserving ${\cal N} =2$ superconformal invariance, thus defining a two-parameter family
of ${\cal N} = 2$ SCFTs. 

A global $SU(2)_L$ symmetry is present for all values of the couplings (apart from  the $SU(2)_R \times U(1)_r$ R-symmetry).
At the orbifold point $g_{YM} = \check g_{YM}$
there is an extra $\mathbb{Z}_2$ symmetry acting as 
\begin{equation} \label{quantumZ2}
\phi \leftrightarrow \check \phi \,, \quad
\lambda_\II \leftrightarrow \check \lambda_\II \, , \quad  A_{\mu} \leftrightarrow \check A_{\mu}  \, ,\quad \psi_{\hat \II} \leftrightarrow \tilde \psi_{\hat \II} \, ,\quad
Q_{\II \hat \II} \leftrightarrow  -\epsilon_{\II\JJ}\epsilon_{\hat{\II}\hat{\JJ}}\bar{Q}^{ \JJ \hat \JJ } \, .
\end{equation}

Setting $\check g_{YM} = 0$, the second vector multiplet $(\check \phi, \check \lambda_\II, \check A_m)$ becomes free and completely decouples 
from the rest of theory, which happens to coincide 
with ${\cal N} = 2$ SCQCD (indeed the field content is the same and ${\cal N} = 2$ susy does the rest). 
The $SU(N_{\check{c}})$
symmetry can now be interpreted as a global flavor symmetry.  What is more,  for $\check g_{YM} = 0$ there is a symmetry enhancement
$SU(N_{\check c}) \times SU(2)_L \to U(N_f = 2 N_c)$ 
 the $SU(N_{\check c})$ index $\check a$
and the $SU(2)_L$ index $\hat \II$
 can  be combined  into a single flavor index $i \equiv (\check a, \hat I) =1, \dots 2 N_c$.

\bibliographystyle{JHEP}

\providecommand{\href}[2]{#2}\begingroup\raggedright\endgroup

\end{document}